\documentclass[aps,prd,twocolumn,superscriptaddress,showpacs]{revtex4}
\usepackage{epsfig,epsf}
\usepackage{amsmath}
\usepackage{amsthm}
\usepackage{amsfonts}
\usepackage{amssymb}
\usepackage{dsfont}
\usepackage{multirow}
\usepackage{appendix}
\usepackage{slashed}
\usepackage[active]{srcltx}
\usepackage{psfrag}

\newcommand{\be}{\begin{equation}}
\newcommand{\ee}{\end{equation}}
\newcommand{\ba}{\begin{eqnarray}}
\newcommand{\ea}{\end{eqnarray}}
\newcommand{\baa}{\begin{eqnarray*}}
\newcommand{\btab}{\begin{tabular}}
\newcommand{\etab}{\end{tabular}}
\newcommand{\eaa}{\end{eqnarray*}}

\def\inbar{\,\vrule height1.5ex width.4pt depth0pt}
\def\IC{\relax\hbox{$\inbar\kern-.3em{\rm C}$}}
\def\IZ{\relax{\hbox{\cmss Z\kern-.4em Z}}}
\def\IR{{\hbox{{\rm I}\kern-.2em\hbox{\rm R}}}}

\def\IP{{\hbox{{\rm I}\kern-.2em\hbox{\rm P}}}}
\def\II{\hbox{{1}\kern-.25em\hbox{l}}}

\begin{document}

\title{ Strong  $D^{*}_sD_{s}\eta^{(\prime)}$ and $%
B^{*}_sB_{s}\eta^{(\prime)}$ vertices from  QCD light-cone sum rules}
\date{\today}
\author{S.~S.~Agaev}
\affiliation{Department of Physics, Kocaeli University, 41380 Izmit, Turkey}
\affiliation{Institute for Physical Problems, Baku State University, Az--1148 Baku,
Azerbaijan}
\author{K.~Azizi}
\affiliation{Department of Physics, Do\v{g}u\c{s} University, Acibadem-Kadik\"{o}y, 34722
Istanbul, Turkey}
\author{H.~Sundu}
\affiliation{Department of Physics, Kocaeli University, 41380 Izmit, Turkey}

\begin{abstract}
The strong  $D^{*}_sD_{s}\eta^{(\prime)}$ and $B^{*}_sB_{s}\eta^{(%
\prime)}$ vertices are studied and the relevant couplings are calculated in the context of
the light-cone  QCD sum rule method with twist-4 accuracy by including
the next-to-leading order corrections. In the
analysis, both the quark and gluon components of the $\eta$ and $\eta^{\prime}$
mesons and the axial anomaly corrected higher twist distributions are included.
\end{abstract}

\maketitle



%

\section{Introduction}

%
During last years the investigation of spectroscopy, electromagnetic, weak and
strong decay channels of heavy mesons, computation of their numerous
transition form factors, strong couplings with different hadrons became one
of the rapidly growing branches of the hadronic physics. The progress in
understanding of the nature of such mesons, including
bottom(charm)-strange ones was achieved from both the experimental and
theoretical sides.

Thus, experimental measurements of hadronic processes and extraction of
parameters of bottom(charm)-strange mesons were performed by different
collaborations \cite{Choi:2015lpc,Aaij:2014bba,Onyisi:2013bjt,Abazov:2012zz,Aaltonen:2012mg, Aubert:2008rs}. Theoretical calculations of parameters related to these mesons were
fulfilled applying various non-perturbative approaches and schemes, such as the lattice QCD calculations \cite{Donald:2013pea}, the QCD and three-point sum rule methods (for instance, see  \cite{Yu:2015xwa,Azizi:2014nta,Sungu:2014lua,Azizi:2013tla,Yazici:2013eia,Rodrigues:2013bta, Khosravi:2013ad,Lucha:2010ea}), and different quark models \cite{Ke:2013zs,Segovia:2009zz}. By this way, masses, strong couplings and form factors of some
bottom(charm)-strange mesons were obtained.

Studies of the vertices consisting of interacting bottom(charm)-strange and
light mesons have also attracted considerable interest. In fact, the strong
couplings determined by the vertices $D^{*}_sD_{s}\eta^{(\prime)}$ and $%
B^{*}_sB_{s}\eta^{(\prime)} $ have been recently calculated in Ref.\cite%
{Yazici:2013eia}, where the three-point sum rule approach has been used.
The present work is devoted to the analysis of these vertices, but within
the context of the QCD light-cone sum rule (LCSR) method \cite{Braun:1989}. The latter provides
more elaborated theoretical tools to perform detailed analysis of
the aforementioned problems. Indeed, the light-cone sum rule method invokes such
quantities of the eta mesons as their distribution amplitudes (DAs) of
different twists and partonic contents. This allows one to take into account
the quark-gluon structure of particles in more clear form than other
approaches.

It should be noted that the $\eta-\eta^{\prime}$ system of light
pseudoscalar mesons accumulate important properties of the particle
phenomenology, like mixing of the $SU(3)$ flavor group singlet $\eta_1$ and
octet $\eta_8$ states to form the physical mesons, the problem of axial $U(1)$
anomaly and its impact on the relevant distribution amplitudes of the eta mesons. To
this list of features one should add also the complicated quark-gluon structure
of the $\eta$ and $\eta^{\prime}$ mesons and subtleties in treatment of their
gluon components that contribute to exclusive processes, the vertices under
consideration being sample ones, at the next-to leading order (NLO) of the
perturbative QCD. These features of the $\eta-\eta^{\prime}$ system,
as well as new experimental data triggered numerous theoretical works
devoted to the analysis of the mesons' mixing problems and computations of various
exclusive processes to extract some constraints on the parameters of their
distributions amplitudes including the two-gluon ones \cite
{Feldmann:1998vh,Agaev:2001rn,Kroll:2002nt,Beneke:2002jn,Agaev:2002ek,Escribano:2005qq,Ball:2007hb, Charng:2006zj,Agaev:2010zz,DiDonato:2011kr,Harland-Lang:2013ncy,Offen:2013nma,Agaev:2014}.
The aim of this work is to study the bottom(charm)-strange
meson strong couplings and consider the vertices $D^{*}_sD_{s}\eta^{(\prime)}$
and $B^{*}_sB_{s}\eta^{(\prime)} $ by including into analysis gluon
component of the $\eta$ and $\eta^{\prime}$ mesons. The computation of a gluonic
contribution to such strong couplings is a new issue that is considered in the present study.

This paper is structured in the following manner. In section \ref{sec:DA},
we present rather comprehensive information on the quark-gluon structure of
the $\eta$ and $\eta^{\prime}$ mesons and details of their leading and
higher twist distribution amplitudes. Existing singlet-octet and
quark-flavor mixing schemes of the $\eta-\eta^{\prime}$ system, their
advantages and drawbacks are briefly outlined. In section \ref{sec:LCSR},
the light-cone sum rules for the strong couplings are derived. Here, the mesons'
leading and higher-twist DAs up to twist-four
are utilized. In this section, we calculate the NLO corrections to the
leading-twist term, and include into the light-cone sum rules also contributions appearing due to
gluon component of the eta mesons. In section \ref{sec:results} we perform
numerical computations to find the values of the corresponding strong couplings. In this section we make
also our brief conclusions. In Appendix A the QCD two-point sum rule expressions
to determine some of parameters in higher twist DAs of the $\eta-
\eta^{\prime}$ system are collected.
\raggedbottom
%

\section{Mixing schemes and distribution amplitudes of $\protect\eta$, $%
\protect\eta^{\prime}$ mesons}

\label{sec:DA} %

Computation of the strong couplings $D^{*}_sD_{s}\eta^{(\prime)}$ and $%
B^{*}_sB_{s}\eta^{(\prime)} $, and relevant matrix elements within the framework
of QCD LCSR method requires knowledge of the $\eta$ and $\eta^{\prime}$
mesons' distribution amplitudes. In this work we use the mixing scheme  for the eta mesons' DAs elaborated in Ref.\ \cite{Agaev:2014} and relevant expressions presented there by adding the necessary formulas for the three-particle twist-3 DAs $\Phi_{3M}^{(s)}(\alpha)$.

Below we concentrate mainly on the $s$-quark
distributions, because only $s$ valence quarks from the heavy $D_{s}^{(\ast)}$
and $B_{s}^{(\ast)}$ mesons contribute to quark-antiquark and quark-gluon-antiquark DAs of the
eta mesons. Nevertheless, when necessary, we provide some information also
on $q$-components of the corresponding DAs.

Hence we define two-particle DAs for the $s$-quark flavor as
\begin{eqnarray}
&&\langle M(q) \mid \overline{s}(x)\gamma _{\mu }\gamma _{5}s(0)\mid 0\rangle
\notag \\
&&=-iq_{\mu }F_{M}^{(s)} \int_{0}^{1}due^{iqxu} \phi_{M}^{(s)}(u,\mu)
\label{eq:LT-DA}
\end{eqnarray}
where $M(q)$ is the $\eta(q)$ or $\eta^{\prime}(q)$ meson state. In this
expression $\phi_{M}^{(s)}(u)$ is the leading twist, i.e. twist-2 DA of the $%
M(q)$ meson. For brevity, in the matrix element, the gauge link is not shown
explicitly. The normalization is chosen such that
\begin{equation}
\int_0^1 du\, \phi_{M}^{(s)}(u,\mu) = 1.\,  \label{eq:Nor}
\end{equation}

The similar distribution amplitudes can be defined for $q=u,\ d$-quarks
as well, with evident replacement $s\to q$ in Eqs.~(\ref{eq:LT-DA}) and (%
\ref{eq:Nor}). Then assuming exact isospin symmetry and denoting
$m_q = (m_u+m_d)/2$
we can determine the couplings $F^{(u)}_M = F^{(d)}_M$, $F^{(s)}_M$ as the
matrix elements
\begin{align}
\langle 0| J^{(i)}_{\mu5}|M(q)\rangle = i f^{(i)}_M q_\mu\,, \,\, i= q,s\,,
\end{align}
of flavor-diagonal axial vector currents $J^{i}_{\mu 5}$
\begin{align}
J^{(q)}_{\mu5} =\frac{1}{\sqrt{2}}\Big[\bar u \gamma_\mu\gamma_5 u + \bar d
\gamma_\mu\gamma_5 d\Big], \,\, J^{(s)}_{\mu5} = \bar s \gamma_\mu\gamma_5
s\,.
\end{align}
The couplings $F^{(u)}_M$, $F^{(d)}_M$ and $F^{(s)}_M$ are connected with $%
f_{M}^{(i)}$ ones by means of the following simple expressions:
\begin{align*}
F^{(u)}_M = F^{(d)}_M = \frac{f^{(q)}_M}{\sqrt{2}}\,, \,\, F^{(s)}_M =
f^{(s)}_M.  \label{eq:bigF}
\end{align*}

This definition of the distributions corresponds to the quark-flavor (QF)
basis introduced to describe mixing in the $\eta$-$\eta^{\prime}$ system. In
QF basis  mixing of the $q$ and $s$ states forms the physical $\eta$ and $%
\eta^{\prime}$ mesons. Alternatively, one can determine DAs of the eta
mesons starting from the singlet-octet (SO) basis of the $SU(3)$ flavor
group. To this end, one introduces the $SU(3)$ flavor-singlet $J^{(1)}_{\mu5}
$ and octet $J^{(8)}_{\mu5}$ currents
\begin{align}
J^{(1)}_{\mu5} =& \frac{1}{\sqrt{3}}\Big[\bar u \gamma_\mu\gamma_5 u + \bar
d \gamma_\mu\gamma_5 d + \bar s \gamma_\mu\gamma_5 s\Big],  \notag \\
J^{(8)}_{\mu5} =& \frac{1}{\sqrt{6}}\Big[\bar u \gamma_\mu\gamma_5 u + \bar
d \gamma_\mu\gamma_5 d -2 \bar s \gamma_\mu\gamma_5 s\Big],
\end{align}
and defines the corresponding matrix elements as
\begin{align}
\langle 0| J^{(i)}_{\mu5}|M(q)\rangle = i f^{(i)}_M q_\mu\,, & & i = 1,8\,.
\end{align}
The eta mesons quark-flavor and singlet-octet combination of the
distributions are connected with each other as,
\begin{align}
\begin{pmatrix}
f_M^{(8)}\phi^{(8)}_M(u,\mu) \\
f_M^{(1)}\phi^{(1)}_M (u,\mu)%
\end{pmatrix}
&= U(\varphi_0)
\begin{pmatrix}
f_M^{(q)}\phi^{(q)}_M(u,\mu) \\
f_M^{(s)}\phi^{(s)}_M(u,\mu)%
\end{pmatrix}%
.
\end{align}
Here
\begin{align}
U(\varphi_0) =
\begin{pmatrix}
\cos\varphi_0 & - \sin\varphi_0 \\
\sin\varphi_0 & \cos\varphi_0%
\end{pmatrix}
=
\begin{pmatrix}
\sqrt{\frac13} & - \sqrt{\frac23} \\
\sqrt{\frac23} & \sqrt{\frac13}%
\end{pmatrix}%
\end{align}
with $\varphi_0 = \arctan(\sqrt{2})$.

In the singlet-octet basis, the scale dependence of the DAs is considerably
simpler than in the QF approach. In fact, SO couplings and DAs do not mix
with each other via renormalization. Moreover, the octet coupling $f_M^{(8)}$
is scale-independent, whereas the singlet coupling $f_M^{(1)}$ evolves due
to the $U(1)$ anomaly \cite{Kodaira:1979pa}:
\begin{align}
f_{M}^{(1)}(\mu )& = f_{M}^{(1)}(\mu_{0})\Big\{ 1+\frac{2 n_f}{\pi \beta_{0}}%
\Big[\alpha_s(\mu)- \alpha_s(\mu_{0})\Big]\Big\},  \label{eq:anomaly}
\end{align}
where $n_f$ is the number of light quark flavors.

This basis is also preferable for solution of the evolution equations. Thus,
the quark-antiquark DAs in the singlet-octet basis can be expanded in terms
of Gegenbauer polynomials $C_{n}^{3/2}(2u-1)$ that are eigenfunctions of the
one-loop flavor-nonsinglet evolution equation:
\begin{equation}
\phi_{M}^{(1,8)}(u,\mu) = 6u\bar u \Big[1+\!\!\sum\limits_{n=2,4,\ldots}\!\!
a_{n,M}^{(1,8)}(\mu) C_{n}^{3/2}(2u-1)\Big].  \label{phiq}
\end{equation}
The sum in Eq.~(\ref{phiq}) runs over polynomials of even dimension $%
n=2,4,\ldots$ implying that the quark-antiquark DAs are symmetric functions
under the interchange of the quark momenta
\begin{equation}
\phi_{M}^{(1,8)}(u,\mu) = \phi_{M}^{(1,8)}(\bar u,\mu)\,.  \label{eq:qDAsym}
\end{equation}
Another twist-2 DA of the $\eta-\eta^{\prime}$ system is connected with its
two-gluon component. This distribution can be defined as non-local matrix
element
\begin{eqnarray}
\lefteqn{\langle M(p)|G_{\mu \nu}(x) \widetilde G^{\mu \nu}(0) \mid 0\rangle
= }  \notag \\
&&{}\hspace*{0.5cm} = \frac{C_F}{2\sqrt{3}}f^{(1)}_M (qx)^2 \int_0^1 du\,
e^{iqxu} \phi_M^{(g)}(u,\mu)\,,  \label{gluonDA}
\end{eqnarray}
where $G_{\mu\nu}=G_{\mu\nu}^{a}\lambda^a/2$ with $tr[\lambda^{a}%
\lambda^{b}]=2\delta^{ab}$. The dual gluon field strength tensor defined as $%
\widetilde G_{\mu\nu} = (1/2)\epsilon_{\mu\nu\alpha\beta}G^{\alpha\beta}$,
and $C_F=4/3$.

The gluon DA is antisymmetric
\begin{equation}
\phi_{M}^{(g)}(u,\mu) = - \phi_{M}^{(g)}(\bar u,\mu)\,
\end{equation}
and can be expanded in a series of Gegenbauer polynomials $%
C_{n-1}^{5/2}(2u-1)$ of odd dimension
\begin{equation}
\phi_{M}^{(g)}(u,\mu) = 30 u^2\bar u^2 \!\!\sum\limits_{n=2,4,\ldots}\!\!
a_{n,M}^{(g)}(\mu)\,C_{n-1}^{5/2}(2u-1)\,.  \label{phig}
\end{equation}
It should be emphasized that the octet components of the eta mesons' DAs are
renormalized multiplicatively to the leading-order and mix with the gluon
components only at the next-to-leading-order, whereas the singlet components mix
with gluon ones already in the LO (see Appendix B in Ref.\ \cite{Agaev:2014}
for details). The values of the parameters $a_{n,M}^{(1,8,g)}$ at a certain
scale $\mu_0$ determine all nonperturbative information on the DAs.

In the exact $SU(3)$ flavor symmetry limit $\eta = \eta_8$, and $%
\eta^{\prime}$ is a flavor--singlet, $\eta^{\prime}= \eta_1$. In this limit $%
f^{(q)}_\eta = f_\pi$ with $f_\pi = 131$~MeV being equal the pion decay
constant. However, it is known empirically that the $SU(3)$-breaking
corrections are large and , as a result, the relation of physical $\eta, \,
\eta^{\prime}$ mesons to the basic octet and singlet states becomes
complicated and involves two different mixing angles, see, e.g., a
discussion in Ref.~\cite{Feldmann:1998vh}.

To avoid these problems and reduce a number of free parameters necessary to
treat the $\eta-\eta^{\prime}$ system, a new mixing scheme (FKS) was
proposed ~\cite{Feldmann:1998vh}. It is used the QF basis and founded on the
observation that vector mesons $\omega$ and $\phi$ are to a very good
approximation pure $\bar u u + \bar d d $ and $\bar s s$ states and the same
is true also for tensor mesons. The smallness of mixing corresponds to the
OZI rule that is phenomenologically very successful. Therefore, if the axial
$U(1)$ anomaly is the only effect that makes the situation in pseudoscalar
channel different, it is natural to suggest that the physical states are related
to the flavor ones by an orthogonal transformation
\begin{align}
\begin{pmatrix}
|\eta\rangle \\
|\eta^{\prime}\rangle%
\end{pmatrix}
= U(\varphi)
\begin{pmatrix}
|\eta_q\rangle \\
|\eta_s\rangle%
\end{pmatrix}%
, & & U(\varphi) =
\begin{pmatrix}
\cos \varphi & - \sin \varphi \\
\sin \varphi & \cos\varphi%
\end{pmatrix}%
.
\end{align}
The assumption on the state mixing implies that the same mixing pattern
applies to the decay constants and to the wave functions, as well. In other
words
\begin{align}
\begin{pmatrix}
f_\eta^{(q)} & f_\eta^{(s)} \\
f_{\eta^{\prime}}^{(q)} & f_{\eta^{\prime}}^{(s)}%
\end{pmatrix}
=& U(\varphi)
\begin{pmatrix}
f_q & 0 \\
0 & f_s%
\end{pmatrix}%
,  \label{eq:QF}
\end{align}
and
\begin{align}
\begin{pmatrix}
f_\eta^{(q)}\phi_\eta^{(q)} & f_\eta^{(s)}\phi_\eta^{(s)} \\
f_{\eta^{\prime}}^{(q)} \phi_{\eta^{\prime}}^{(q)} & f_{\eta^{\prime}}^{(s)}%
\phi_{\eta^{\prime}}^{(s)}%
\end{pmatrix}
= U(\varphi)
\begin{pmatrix}
f_q\phi_q & 0 \\
0 & f_s\phi_s%
\end{pmatrix}%
,  \label{eq:QFt2}
\end{align}
are held with the same mixing angle $\varphi$.

This conjecture allows one to reduce four DAs of physical states $%
\eta,\eta^{\prime}$ to the two DAs, $\phi_q(u,\mu)$ and $\phi_s(u,\mu)$ of the
flavor states:
\begin{align}
&\phi_\eta^{(q)}(u) = \phi_{\eta^{\prime}}^{(q)}(u) = \phi_q(u)\,,  \notag \\
&\phi_\eta^{(s)}(u) = \phi_{\eta^{\prime}}^{(s)}(u) = \phi_s(u)\,.
\label{eq:QFmodel1}
\end{align}
The singlet and octet DAs is this scheme are given by
\begin{align}
\begin{pmatrix}
f_\eta^{(8)}\phi_\eta^{(8)} & f_\eta^{(1)}\phi_\eta^{(1)} \\
f_{\eta^{\prime}}^{(8)}\phi_{\eta^{\prime}}^{(8)} & f_{\eta^{\prime}}^{(1)}%
\phi_{\eta^{\prime}}^{(1)}%
\end{pmatrix}
=& U(\varphi)
\begin{pmatrix}
f_q\phi_q & 0 \\
0 & f_s\phi_s%
\end{pmatrix}
U^T(\varphi_0)  \label{relation}
\end{align}
and the same relation is valid for the couplings $f^{(i)}_M$ and the
couplings multiplied by the parameters $f^{(i)}_M a^{(i)}_{n,M}$.
The couplings $f_q$ and $f_s$, as well as mixing angle $\varphi$ in the quark-flavor
scheme have been determined in Ref.~\cite{Feldmann:1998vh} from the fit to
the experimental data
\begin{align}
f_q =& (1.07\pm 0.02)f_\pi\,,  \notag \\
f_s =& (1.34\pm 0.06)f_\pi\,,  \notag \\
\varphi =& 39.3^{\circ}\pm 1.0^\circ.  \label{FKSvalues}
\end{align}
It is worth noting that the flavor-singlet and flavor-octet couplings have
different scale dependence, and Eq.~(\ref{relation}) cannot hold at all
scales. It is natural to assume that the scheme refers to a low
renormalization scale $\mu_0 \sim 1$~GeV and the DAs at higher scales are
obtained by the QCD evolution.

Then for the gluon DA we assume that
\begin{equation*}
\langle \eta_{q}|G_{\mu \nu}(x) \widetilde G^{\mu \nu}(0) \mid 0\rangle
=\langle \eta_{s}|G_{\mu \nu}(x) \widetilde G^{\mu \nu}(0) \mid 0\rangle
\end{equation*}
and as a result get:
\begin{align}
\phi_\eta^{(g)}(u) = \phi_{\eta^{\prime}}^{(g)}(u).\,  \label{eq:QFmodel2}
\end{align}

We define two-particle twist-3 DAs for the strange quarks in the following
way
\begin{equation}
2m_{s}\langle M(q)\mid \overline{s}(x)i\gamma _{5}s(0)\mid 0\rangle
=\int_{0}^{1}due^{iqxu}\phi _{3M}^{(s)p}(u)  \label{TW3:ps}
\end{equation}
and
\begin{eqnarray}
2m_{s}\langle M(q)\mid \overline{s}(x)\sigma _{\mu \nu }\gamma_{5}s(0)\mid 0\rangle
\notag \\
=\frac{i}{6}\left( q_{\mu }x_{\nu }-q_{\nu }x_{\mu }\right)
\int_{0}^{1}due^{iqxu}\phi _{3M}^{(s)\sigma }(u)
\label{TW3:sigma}
\end{eqnarray}
with the normalization
\begin{equation}
\int_0^1 du\, \phi_{3M}^{(s)p}(u) = \int_0^1 du\, \phi_{3M}^{(s)\sigma}(u) =
h_{M}^{(s)}.\,  \label{norm3}
\end{equation}
Here \cite{Beneke:2002jn,Agaev:2014}
\begin{align}
h_{M}^{(s)} &= m_{M}^2 f_{M}^{(s)} - A_{M}\,,  \notag \\
A_M &= \langle 0 | \frac{\alpha_s}{4\pi}G^a_{\mu\nu}\widetilde
G^{a,\mu\nu}|M(p)\rangle\,,  \label{eq:aM}
\end{align}
that follows from the anomaly relation
\begin{equation*}
\partial^\mu J^{(s)}_{\mu5} = 2m_s \bar s i\gamma_5 s + \frac{\alpha_s}{4\pi}%
G^a_{\mu\nu}\widetilde G^{a,\mu\nu}.  \label{anomaly}
\end{equation*}
Twist-3 DAs for the light $q=u$ or $d$ quark can be defined by similar
expressions with substitutions $s\to q$, e.g. $H_{M}^{(q)} = m_{M}^2
F_{M}^{(q)} - A_{M}$, where
\begin{align}
H^{(u)}_M = H^{(d)}_M = \frac{h^{(q)}_M}{\sqrt{2}}.  \label{eq:bigH}
\end{align}
Writing the normalization of the twist-3 DAs in this form (see Eqs.\ (\ref{TW3:ps})-(\ref{eq:aM})) we
follow Refs. \cite{Ball:2007hb,Beneke:2002jn,Agaev:2014}. Note that this definition formally remains correct in the chiral $m_s \to 0$ limit. As mentioned above, in this case $\eta$ and $\eta^{\prime}$ are purely flavor-octet and 
flavor-singlet, respectively, so that $\eta$ becomes massless and $\eta^{\prime}$ remains massive
due to the axial anomaly \cite{Witten:1979,Veneziano:1979}. Equation  (\ref{eq:aM}) is then satisfied trivially for
$\eta$, because all three terms vanish, and for $\eta^{\prime}$ the cancelation of the two terms on the r.h.s.
implies the well-known relation for the $\eta^{\prime}$ mass in terms of the anomaly matrix element. The
ratio $h_s/m_s$ (and the similar ratios for light quarks) remains finite so that the contribution of twist-three
DAs to correlation functions remains finite in the case that they enter the coefficients without a quark mass factor. For further discussion and examples we refer to the work \cite{Ball:2007hb}.

We assume that at low scales the FKS mixing scheme is valid for all
quantities and distributions, and introduce two new parameters $h_q$ and $h_s
$~\cite{Beneke:2002jn}
\begin{equation}
\left(
\begin{array}{c}
h_{\eta }^{(q)},\ \ h_{\eta }^{(s)} \\
h_{\eta ^{\prime }}^{(q)},\ \ h_{\eta ^{\prime }}^{(s)}%
\end{array}%
\right) =U(\varphi )\left(
\begin{array}{c}
h_{q},\ \ 0 \\
0,\ \ h_{s}%
\end{array}%
\right)
\end{equation}
with numerical values (in $\mathrm{{GeV}^3}$)
\begin{align}
h_q = 0.0016\pm 0.004\,, & & h_s = 0.087\pm 0.006\,.  \label{hvalues}
\end{align}

Within the FKS scheme, we can rewrite four DAs $\phi_{3M}^{(q,s) p
}$ in terms of two functions $\phi_{3s}^{p}(u)$ and $\phi_{3q}^{p}(u)$. The
same argumentation is valid for the distribution amplitudes $%
\phi_{3M}^{(q,s)\sigma}$, as well. Let us note that for calculation of the
strong couplings of interest we need only $s$-components of the DAs.
Therefore, we get:
\begin{align}
\phi_{3\eta^{\prime}}^{(s)p}(u) = \phi_{3s}^{p}(u)\cos\varphi \,\,, & &
\phi_{3\eta}^{(s)p}(u) = - \phi_{3s}^{p}(u)\sin\varphi \,\,,  \notag \\
\phi_{3\eta^{\prime}}^{(s)\sigma}(u) = \phi_{3s}^{\sigma}(u)\cos\varphi
\,\,, & & \phi_{3\eta}^{(s)\sigma}(u) = - \phi_{3s}^{\sigma}(u)\sin\varphi
\,\,,
\end{align}
where
\begin{align}
\phi_{3s}^{p}(u) &= h_s + 60 m_s f_{3s} C^{1/2}_2(2u-1),  \notag \\
\phi_{3s}^{\sigma}(u) &= 6\bar u u \Big[h_s + 10 m_s f_{3s} C^{3/2}_2(2u-1)
\Big].
\label{eq:tw3das}
\end{align}
The coupling $f_{3s}$ is defined as
\begin{align*}
\langle 0 |\bar s \sigma_{n\xi}\gamma_5 g G^{n \xi} s | \eta_s (p) \rangle
&= 2i (pz)^2 f_{3s},
\end{align*}
and we assume that
\begin{equation}
f^{(s)}_{3\eta^{\prime}} =f_{3s} \cos\varphi \,,\,\, f^{(s)}_{3\eta} = -
f_{3s}\sin\varphi.  \label{eq:const}
\end{equation}
For the coupling $f_{3s}$, as an estimate, we adopt a value of the similar
parameter obtained for the pion. The latter at the scale $\mu_0=1\,\,
\mathrm{GeV}$ is equal to
\begin{equation*}
f_{3s}(\mu_0) \simeq f_{3\pi}(\mu_0)=(0.0045 \pm 0.0015)~\text{GeV}^2.
\end{equation*}
The scale dependence of $f_{3s}(\mu)$ is determined by formula
\begin{equation}
f_{3s}(\mu) = \left[ \frac{\alpha_s(\mu)}{\alpha_s(\mu_0)}\right
]^{55/9\beta_0} f_{3s}(\mu_0).
\end{equation}

Here some comments are in order. Let us explain our choice
of the parameters in the higher-twist DAs. First of all, there is not any
information on flavor-singlet contributions to these parameters. Moreover,
computation of these parameters using the QCD sum rule method by taking into
account only quark contents of $\eta$ and ${\eta^{\prime}}$ mesons lead to numerical
values that are very close to parameters of the pion DAs. In
fact, calculations of the parameters $f_{3s}$ and $\delta_{M}^{2(s)}$ presented
in Appendix A illustrate correctness of such choice. Therefore, in what follows
we will use parameters from the pion DAs keeping in mind that the approximation
accepted here does not encompass the flavor-singlet effects.

The eta mesons' three-particle twist-3 DAs are defined in accordance with
Ref.\ \cite{Braun:1989iv}
\begin{eqnarray}
&&\langle M(q)\mid \overline{s}(x)gG_{\mu \nu }(vx)\sigma _{\alpha \beta
}\gamma _{5}s(0)\mid 0\rangle =  \notag \\
&&if_{3M}^{(s)}\left[ q_{\alpha }\left( q_{\mu }g_{\nu \beta }-q_{\nu
}g_{\mu \beta }\right) -\left (\alpha \leftrightarrow \beta \right )\right ]  \notag \\
&&\times \int D\underline{\alpha }e^{iqx(\alpha _{1}+v\alpha
_{3})}\Phi _{3M}^{(s)}(\alpha),
\end{eqnarray}
where
\begin{align*}
\int\mathcal{D}\underline{\alpha} &= \int_0^1 d\alpha_1 d\alpha_2 d\alpha_3
\delta\big(1-\sum \alpha_i\big).
\end{align*}
The expansion of the function $\Phi_{3M}^{(s)}(\alpha)$ in the conformal spin
leads to the known expression
\begin{equation}
\Phi_{3M}^{(s)}(\alpha)=360\alpha_1\alpha_2\alpha_3^2\left [1+\frac{1}{7}%
\omega_{3s}\left( 7\alpha_3-3\right) \right ],
\end{equation}
with
\begin{equation}
\omega_{3s}(\mu_0)\simeq \omega_{3\pi}(\mu_0)=(-1.5\pm 0.7)\,\, \mathrm{GeV}%
^2,
\end{equation}
and
\begin{equation*}
\left (f_{3s}\omega_{3s}\right )(\mu) = \left[ \frac{\alpha_s(\mu)}{%
\alpha_s(\mu_0)}\right ]^{104/9\beta_0} \left (f_{3s}\omega_{3s}\right
)(\mu_0).
\end{equation*}
Finally, we will need the DAs of twist-4 that are rather numerous. First of
all, there are four two-particle twist-4 distribution amplitudes of the $%
\eta-\eta^{\prime}$ system stemming from the matrix element
\begin{eqnarray}
\langle M(q) \mid \overline{s}(x)\gamma _{\mu }\gamma _{5}s(0)\mid 0\rangle
=-iq_{\mu }F_{M}^{(s)}  \notag \\
\times \int_{0}^{1}due^{iqxu}\left[ \phi _{M}^{(s)}(u)+\frac{x^{2}}{16}\phi
_{4M}^{(s)}(u)\right]  \notag \\
-i\frac{x_{\mu }}{qx}F_{M}^{(s)}\int_{0}^{1}due^{iqxu}\psi _{4M}^{(s)}(u).
\end{eqnarray}
Another three-particle twist-4 distributions are given by the expressions:
\begin{eqnarray}
\langle M(q)\mid \overline{s}(x)\gamma _{\mu }\gamma _{5}g_{s}G_{\alpha
\beta }(vx)s(0)\mid 0\rangle =  \notag \\
F_{M}^{(s)}\frac{q_{\mu }}{qx}\left( q_{\alpha }x_{\beta }-q_{\beta
}x_{\alpha }\right) \int \mathcal{D}\underline{\alpha}e^{iqx(\alpha
_{1}+v\alpha _{3})}\Phi _{4M}^{(s)}(\alpha)  \notag \\
+F_{M}^{(s)}\left[ q_{\beta }\left(g_{\alpha \mu }-\frac{x_{\alpha }q_{\mu }%
}{qx}\right) -q_{\alpha }\left( g_{\beta \mu }-\frac{x_{\beta }q_{\mu }}{qx}%
\right) \right]  \notag \\
\times \int \mathcal{D}\underline{\alpha }e^{iqx(\alpha _{1}+v\alpha
_{3})}\Psi _{4M}^{(s)}(\alpha),
\end{eqnarray}
and
\begin{eqnarray}
\langle M(q)\mid \overline{s}(x)\gamma _{\mu }\gamma _{5}g_{s}\widetilde{G}%
_{\alpha \beta }(vx)s(0)\mid 0\rangle =  \notag \\
F_{M}^{(s)}\frac{q_{\mu }}{qx}\left( q_{\alpha }x_{\beta }-q_{\beta
}x_{\alpha }\right) \int \mathcal{D} \underline{\alpha }e^{iqx(\alpha
_{1}+v\alpha _{3})}\widetilde{\Phi} _{4M}^{(s)}(\alpha)  \notag \\
+F_{M}^{(s)}\left[ q_{\beta }\left(g_{\alpha \mu }-\frac{x_{\alpha }q_{\mu }%
}{qx}\right) -q_{\alpha }\left( g_{\beta \mu }-\frac{x_{\beta }q_{\mu }}{qx}%
\right) \right]  \notag \\
\times \int \mathcal{D} \underline{\alpha }e^{iqx(\alpha _{1}+v\alpha _{3})}%
\widetilde{\Psi} _{4M}^{(s)}(\alpha).
\end{eqnarray}

The distribution amplitudes $\Phi_{4M}^{(s)}$ and $\Psi_{4M}^{(s)}$ can be
expanded in orthogonal polynomials that correspond to contributions of
increasing spin in the conformal expansion. Taking into account contributions
of the lowest and the next-to-lowest spin
one finds~\cite{Braun:1989iv,Ball:1998je,Ball:2006wn,Agaev:2014}
\begin{eqnarray}
\Phi_{4M}^{(s)}(\alpha) & = & 120 \alpha_1\alpha_2\alpha_3 \Big[%
\phi_{1,M}^{(s)} (\alpha_1-\alpha_2)\Big],  \notag \\
\widetilde\Phi_{4M}^{(s)}(\alpha) & = & 120 \alpha_1\alpha_2\alpha_3 \Big[ %
\widetilde\phi_{0,M}^{(s)} + \widetilde\phi_{2,M}^{(s)} (3\alpha_3-1)\Big],
\notag \\
{\widetilde\Psi}^{(s)}_{4M}(\alpha) & = & -30 \alpha_3^2\Big\{ %
\psi^{(s)}_{0,M}(1-\alpha_3)  \notag \\
&&{} \hspace*{0.8cm} +\psi^{(s)}_{1,M}\Big[\alpha_3(1\!-\!\alpha_3)-6%
\alpha_1\alpha_2\Big]  \notag \\
&&{} \hspace*{0.8cm} +\psi^{(s)}_{2,M}\Big[\alpha_3(1\!-\!\alpha_3)-\frac{3}{%
2}(\alpha_1^2 +\alpha_2^2)\Big]\Big\},  \notag \\
{\Psi}^{(s)}_{4M}(\alpha) & = & - 30 \alpha_3^2 (\alpha_1-\alpha_2) \Big\{ %
\psi^{(s)}_{0,M} + \psi^{(s)}_{1,M} \alpha_3  \notag \\
&&{} \hspace*{2.5cm} + \frac{1}{2} \psi^{(s)}_{2,M}(5 \alpha_3-3)\Big\}.
\label{eq:T4-conformal}
\end{eqnarray}
The coefficients $\phi_{kM}^{(s)}$, $\psi_{kM}^{(s)}$ are related by QCD
equations of motion (EOM)~\cite{Agaev:2014}. From these EOM one obtains
\begin{align}
\widetilde\phi_{0M}^{(s)} = \psi^{(s)}_{0M} = - \frac13\,
\delta^{2(s)}_{M}\,,
\end{align}
and
\begin{align}
\widetilde\phi_{2M}^{(s)} &= \frac{21}{8} \delta^{2(s)}_{M}
\omega_{4M}^{(s)},  \notag \\
\phi_{1M}^{(s)} &= \frac{21}{8}\left[\delta^{2(s)}_{M}\omega_{4M}^{(s)} +
\frac{2}{45} m^2_M \left(1-\frac{18}{7}a^{(s)}_{2M}\right) \right],  \notag
\\
\psi_{1M}^{(s)} & = \frac{7}{4} \left[ \delta^{2(s)}_{M}\omega_{4M}^{(s)}
\!+\! \frac{1}{45} m^2_M \left(1\!-\!\frac{18}{7}a^{(s)}_{2M}\right) \!+\!
4m_s \frac{f_{3M}^{(s)}}{f^{(s)}_M} \right]\!,  \notag \\
\psi_{2M}^{(s)} & = \frac{7}{4} \left[ 2 \delta^{2(s)}_{M}\omega_{4M}^{(s)}
\!-\! \frac{1}{45} m^2_M \left(1\!-\!\frac{18}{7}a^{(s)}_{2M}\right) \!-\! 4
m_s \frac{f_{3M}^{(s)}}{f^{(s)}_M} \right]\!.  \label{NLOspin}
\end{align}
Here the parameter $\delta^{2(s)}_{M}$ is defined as
\begin{align*}
\langle 0| \bar s \gamma^\rho ig \widetilde{G}_{\rho\mu} s |M(p)\rangle =
p_\mu f_M^{(s)} \delta^{2(s)}_{M}.\,
\end{align*}
Its value at $\mu_0$ is chosen equal to
\begin{equation}
\delta_{M}^{2(s)}(\mu_0)\simeq \delta_{\pi}^2(\mu_0)=(0.18\pm 0.06) \,\,%
\mathrm{GeV}^2,
\end{equation}
and evolution is given by the formula
\begin{equation*}
\delta_{M}^{2(s)}(\mu) = \left[ \frac{\alpha_s(\mu)}{\alpha_s(\mu_0)}\right
]^{10/\beta_0}\delta_{M}^{2(s)}(\mu_0).
\end{equation*}
We set the parameter $\omega_{4M}^{(s)}(\mu_0)$ equal to $%
\omega_{4\pi}(\mu_0)$:
\begin{equation}
\omega_{4M}^{(s)}(\mu_0) \simeq \omega_{4\pi}(\mu_0)=(0.2\pm 0.1) \,\,
\mathrm{GeV}^2,
\end{equation}
with
\begin{equation*}
\left(\delta_{M}^{2(s)}\omega_{4M}^{(s)}\right )(\mu) = \left[ \frac{%
\alpha_s(\mu)}{\alpha_s(\mu_0)}\right
]^{32/9\beta_0}\left(\delta_{M}^{2(s)}\omega_{4M}^{(s)}\right )(\mu_0).
\end{equation*}

The DAs $\phi^{(s)}_{4M}(u)$ and $\psi^{(s)}_{4M}(u)$ can be calculated in
terms of the three-particle DAs of twist four and the DAs of lower twist. As
a result, one obtains the expressions for the two-particle DAs $%
\psi_{4M}^{(s)}(u)$ and $\psi_{4M}^{(s)}(u)$ that can be separated in
``genuine'' twist-four contributions and meson mass corrections as
\begin{align}
\psi_{4M}^{(s)}(u) = \psi_{4M}^{(s)\mathrm{twist}}(u) + m^2_M \psi_{4M}^{(s)%
\mathrm{mass}}(u)
\end{align}
with
\begin{eqnarray}
\psi_{4M}^{(s)\mathrm{twist}}(u) &=& \frac{20}{3} \delta^{2(s)}_M
C_2^{1/2}(2u-1) + 30 m_s \frac{f^{(s)}_{3M}}{f^{(s)}_{M}}  \notag \\
&&{} \times \Big(\frac{1}{2}-10 u\bar u +35 u^2\bar u^2\Big)\,,  \notag \\
\psi_{4M}^{(s)\mathrm{mass}}(u) &=& \frac{17}{12} - 19 u\bar u + \frac{105}{2%
} u^2\bar u^2  \notag \\
&&{} + a^{(s)}_{2,M} \Big(\frac{3}{2} - 54 u\bar u + 225 u^2\bar u^2 \Big)
\end{eqnarray}
and similarly
\begin{align}
\phi_{4M}^{(s)}(u) = \phi_{4M}^{(s)\mathrm{twist}}(u) + m^2_M \phi_{4M}^{(s)%
\mathrm{mass}}(u)\,,
\end{align}
where
\begin{eqnarray}
\phi_{4M}^{(s)\mathrm{twist}}(u) &=& \frac{200}{3}\delta^{2(s)}_M u^2 \bar
u^2 + 21 \delta^{2(s)}_M \omega_{4M}^{(s)}\Big\{ u\bar u (2\!+\!13 u\bar u)
\notag \\
&&{} + 2\big[ u^3(10-15 u+6 u^2)\ln u + (u\leftrightarrow \bar u)\big]\Big\}
\notag \\
&&{} + 20 m_s \frac{f^{(s)}_{3M}}{f^{(s)}_{M}} u\bar u \Big[12 -63u \bar u +
14u^2 \bar u^2\Big],  \notag \\
\phi_{4M}^{(s)\mathrm{mass}}(u) &=& u\bar u \Big[\frac{88}{15} + \frac{39}{5}
u\bar u + 14 u^2\bar u^2\Big]  \notag \\
&&{} - a^{(s)}_{2,M} u\bar u \Big[ \frac{24}{5} - \frac{54}{5} u\bar u + 180
u^2\bar u^2\Big]  \notag \\
&& + \Big(\frac{28}{15}-\frac{24}{5} a^{(s)}_{2,M} \Big)\Big[ u^3(10-15 u+6
u^2)\ln u  \notag \\
&&{} + (u\leftrightarrow \bar u)\Big].
\end{eqnarray}
These expressions complete the list of the distribution amplitudes that are
necessary for analyzing the strong vertices $D^{*}_sD_{s}\eta^{(\prime)}$
and $B^{*}_sB_{s}\eta^{(\prime)}$ with twist-4 accuracy.

It is worth noting that we have chosen parameters of the higher-twist DAs in
order to obey pattern of the state mixing accepted for the $%
\eta-\eta^{\prime}$ system. In fact, it is not difficult to see that
relations in Eq.\ (\ref{eq:QFt2}) are true for DAs $\phi_{3M}^{(3)p}(u)$, $%
\phi_{3M}^{(3)\sigma}(u)$ and $f_{3M}^{(s)}\Phi_{3M}^{(s)}(\alpha)$, as
well. This formula are fulfilled approximately for twist-4 DAs $%
F_{M}^{(s)}\phi_{4M}^{(s)}(u)$ and $F_{M}^{(s)}\psi_{4M}^{(s)}(u)$. The main
sources of deviation from Eq.\ (\ref{eq:QFt2}) are terms $\sim m_{M}^2$ in
twist-4 DAs that, nevertheless, numerically have rather small effects on
final results.

%

\section{The LCSR for strong couplings}

\label{sec:LCSR} %
In the context of the QCD sum rules on the light-cone heavy-heavy-light
meson strong couplings were analyzed already in Refs. \cite%
{Braun:1995,Aliev:1996,Khodjamirian:1999}, where the vertices $D^{\ast }D\pi
$, $B^{\ast }B\pi $, as well as vertices with $\rho $-meson were considered.
In the present work we calculate within the QCD LCSR method the strong couplings
that correspond to the vertices $D_{s}^{\ast }D_{s}\eta^{(\prime )}$ and
$B_{s}^{\ast }B_{s}\eta ^{(\prime )}$. Below we
concentrate on the couplings $g_{B_{s}^{\ast }B_{s}M}$: results for $%
g_{D_{s}^{\ast }D_{s}M}$ can be easily obtained from relevant expressions by
replacements $b\rightarrow c$, $B_{s}^{0}\rightarrow D_{s}^{-}$ and $%
B_{s}^{0\ast }\rightarrow D_{s}^{\ast -}$.
%

\subsection{Leading order results}

%
In calculation of the leading order contribution to the LCSR we use technical tools and methods elaborated
in the original paper \cite{Braun:1995}.
We start from the
correlation function
\begin{eqnarray}
F_{\mu }(p,q)&=&i\int d^{4}xe^{ipx}\langle M(q)\mid \mathcal{T}\left\{
\overline{s}(x)\gamma _{\mu }b(x)\right.,  \notag \\
&&\left. \overline{b}(0)i\gamma _{5}s(0)\right\} \mid 0\rangle .
\label{eq:Corr}
\end{eqnarray}
It is well known that this correlator can be calculated in both hadronic and
quark-gluon degrees of freedom. Within the QCD LCSR method obtained by this
way expressions should be matched in order to find the couplings $g_{B_{s}^{\ast
}B_{s}\eta}$ and $g_{B_{s}^{\ast }B_{s}\eta^{\prime}}$, and extract
numerical estimates for them.
\begin{figure}[t]
\centerline{
\begin{picture}(210,140)(0,0)
\put(-5,0){\epsfxsize7.8cm\epsffile{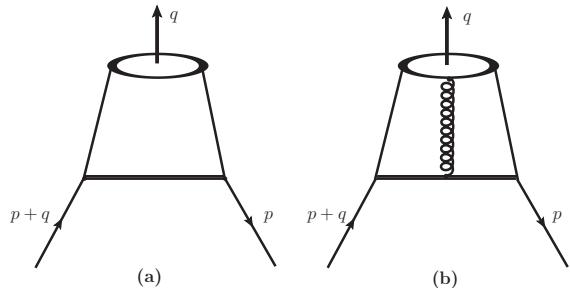}}
\end{picture}
}
\caption{Leading order diagrams contributing to the correlation function.
Thick lines correspond to a heavy quark. Diagram (a) describes
quark-antiquark contributions of various twists to the correlator, whereas
(b) is contribution coming from three-particle components of the meson
distribution amplitude.}
\label{fig:LCSR-1}
\end{figure}
In terms of hadronic quantities, the aforementioned correlation functions are
given by the expression
\begin{eqnarray*}
F_{\mu }^{h}(p,q) &=&\frac{g_{B_{s}^{\ast
}B_{s}M}m_{B_{s}}^{2}m_{B_{s}^{\ast }}f_{B_{s}}f_{B_{s}^{\ast }}}{%
m_{b}\left( p^{2}-m_{B_{s}^{\ast }}^{2}\right) \left[ \left( p+q\right)
^{2}-m_{B_{s}}^{2}\right] } \\
&&\times \left[ q_{\mu }+\frac{1}{2}\left( 1-\frac{m_{B_{s}}^{2}+m_{M}^{2}}{%
m_{B_{s}^{\ast }}^{2}}\right) p_{\mu }\right] ,
\end{eqnarray*}%
where we have defined the couplings $g_{B_{s}^{\ast }B_{s}M}$ and decay
constants $f_{B_s}$, $f_{B_{s}^{*}}$ by means of the following matrix
elements:
\begin{eqnarray}
&&\langle B_{s}^{\ast 0}(p)M(q) \mid B_{s}^{0}(p+q)\rangle =-g_{B_{s}^{\ast
}B_{s}M}q_{\mu }\epsilon ^{\mu },  \notag \\
&&\langle B_{s} \mid \overline{b}i\gamma _{5}s\mid 0\rangle =\frac{m_{
B_{s}}^{2}f_{B_{s}}}{m_{b}},  \notag \\
&&\langle 0 \mid \overline{s}\gamma _{\mu }b\mid B_{s}^{\ast }\rangle
=m_{B_{s}^{\ast }}f_{B_{s}^{\ast }}\epsilon _{\mu }.
\end{eqnarray}
The correlation function depends on the invariants $p^2$, $(p+q)^2$, and can
be written as a sum of invariant amplitudes
\begin{equation*}
F_{\mu}(p,q)=F(p^2,(p+q)^2)q_{\mu}+{\widetilde F}(p^2,(p+q)^2)p_{\mu}.
\end{equation*}
For our purposes it is enough to consider the function $F(p^2,(p+q)^2)$.

Computation of the amplitude $F(p^{2},(p+q)^{2})$ in terms of the hadronic
quantities leads to expression that contains contribution of the ground
state, and a contribution of the higher resonances and continuum states with
relevant quantum numbers in a form of double dispersion integral
\begin{eqnarray}
&&F^{h}(p^{2},(p+q)^{2})=\frac{g_{B_{s}^{\ast
}B_{s}M}m_{B_{s}}^{2}m_{B_{s}^{\ast }}f_{B_{s}}f_{B_{s}^{\ast }}}{%
m_{b}\left( p^{2}-m_{B_{s}^{\ast }}^{2}\right) \left[ \left( p+q\right)
^{2}-m_{B_{s}}^{2}\right] }  \notag \\
&&+\int \frac{ds_{1}ds_{2}\rho ^{h}(s_{1},s_{2})}{%
(s_{1}-p^{2})[s_{2}-(p+q)^{2}]} +\ldots
\end{eqnarray}%
Here the dots stand for single dispersion integrals that, in general,
should be included to make the expression finite.

Having considered $p^{2}$ and $(p+q)^{2}$ as independent variables and
applied the Borel transformation we find
\begin{eqnarray}
&&\mathcal{B}_{M_{1}^{2}}\mathcal{B}_{M_{2}^{2}}F^{h}(p^{2},\,(p+q)^{2})=%
\frac{g_{B_{s}^{\ast }B_{s}M}m_{B_{s}}^{2}m_{B_{s}^{\ast
}}f_{B_{s}}f_{B_{s}^{\ast }}}{m_{b}}  \notag \\
&&\times e^{-\frac{m_{B_{s}^{\ast }}^{2}}{M_{1}^{2}}-\frac{m_{B_{s}^{\ast
}}^{2}}{M_{2}^{2}}}+\int ds_{1}ds_{2}e^{-\frac{s_{1}}{M_{1}^{2}}-\frac{s_{2}%
}{M_{2}^{2}}}\rho ^{h}(s_{1},\,s_{2}).  \label{eq:BHad}
\end{eqnarray}

In order to obtain sum rules expression for the strong couplings, the double
Borel transformation should be applied to the same invariant amplitude, but
now calculated using the quark-gluon degrees of freedom. To this end, one
needs to employ the general expression for the correlation function Eq.\ (\ref%
{eq:Corr}) and compute it by substituting the light-cone expansion for the $b
$-quark propagator
\begin{eqnarray}
&\langle &0\mid \mathcal{T}\{b(x)\overline{b}(0)\}\mid 0\rangle =\int \frac{%
d^{4}k}{(2\pi )^{4}i}e^{-ikx}\frac{{\slashed k}+m_{b}}{m_{b}^{2}-k^{2}}
\notag \\
&&-ig_{s}\int \frac{d^{4}k}{(2\pi )^{4}}e^{-ikx}\int_{0}^{1}dv\left[ \frac{1%
}{2}\frac{{\slashed k}+m_{b}}{\left( m_{b}^{2}-k^{2}\right) ^{2}}G^{\mu \nu
}\left( vx\right) \sigma _{\mu \nu }\right.  \notag \\
&&\left. +\frac{{\slashed k}+m_{b}}{m_{b}^{2}-k^{2}}vx_{\mu }G^{\mu \nu
}\left( vx\right) \gamma _{\nu }\right],  \label{eq:Cprop}
\end{eqnarray}%
and expressing remaining non-local matrix elements in terms of distribution
amplitudes of the eta mesons. The diagrams corresponding to the free $b$%
-quark propagator, and to the one-gluon field components in the expansion Eq.\ (%
\ref{eq:Cprop}) are depicted in Fig.\ \ref{fig:LCSR-1}(a) and Fig.\ \ref%
{fig:LCSR-1}(b), respectively.

Technical details of similar calculations can be found in Ref.\ \cite%
{Braun:1995}. Therefore, we do not concentrate here on these procedures and
provide below only final results. Thus, for the contribution arising from
the diagram (a) we find
\begin{eqnarray}
&&F^{(a)}(p^{2},(p+q)^{2})=\int_{0}^{1}\frac{du}{\Delta (p,q,u)}\left\{
m_{b}F_{M}^{(s)}\left[ \phi _{M}^{(s)}(u)\right. \right.  \notag \\
&&\left. \left. -\frac{m_{M}^{2}u\overline{u}}{\Delta (p,q,u)}\phi
_{M}^{(s)}(u)+\frac{1}{\Delta (p,q,u)}\left( 2uG_{4M}^{(s)}(u)\right.
\right. \right.  \notag \\
&&\left. \left. \left. -\frac{m_{b}^{2}\phi _{4M}^{(s)}(u)}{2\Delta (p,q,u)}%
\right) \right] + \frac{\phi _{3M}^{(s)p}(u)}{2m_{s}}u+\frac{\phi
_{3M}^{(s)\sigma }(u)}{6m_{s}} \right.  \notag \\
&&\left. \left. +\frac{\phi _{3M}^{(s)\sigma }(u)}{12m_{s}}\frac{%
m_{b}^{2}+p^{2}}{\Delta (p,q,u)}\right. \right\} .  \label{eq:fa}
\end{eqnarray}%
In this expression we have introduced the short-hand notation for the
denominator of the free $b$-quark propagator (see first term in Eq.\ ({\ref%
{eq:Cprop}}))
\begin{equation*}
\Delta (p,q,u)=m_{b}^{2}-(1-u)p^{2}-u(p+q)^{2}
\end{equation*}%
and also defined the new function $G_{4M}^{(s)}(u)$
\begin{equation*}
G_{4M}^{(s)}(u)=-\int_{0}^{u}\psi _{4M}^{(s)}(v)dv.
\end{equation*}%
The meson mass correction $\sim m_{M}^2$ in Eq.\ (\ref{eq:fa}) comes from
the expansion of the leading order twist-2 term.

Computations with one-gluon field components in the $b$-quark propagator
lead to the following result:
\begin{eqnarray}
&&F^{(b)}(p^{2},(p+q)^{2})=\int_{0}^{1}dv\int \mathcal{D}\alpha  \notag \\
&&\times \left\{ \frac{4f_{3M}^{(s)}\Phi _{3M}^{(s)}(\alpha )vpq}{\left[
m_{b}^{2}-\left( p+q\left( \alpha _{1}+v\alpha _{3}\right) \right) ^{2}%
\right] ^{2}}\right.  \notag \\
&&\left. +F_{M}^{(s)}m_{b}\frac{2\Psi _{4M}^{(s)}(\alpha )-\Phi
_{4M}^{(s)}(\alpha )+2\widetilde{\Psi }_{4M}^{(s)}(\alpha )-\widetilde{\Phi }%
_{4M}^{(s)}(\alpha )}{\left[ m_{b}^{2}-\left( p+q\left( \alpha _{1}+v\alpha
_{3}\right) \right) ^{2}\right] ^{2}}\right\}  \notag \\
&&{}
\end{eqnarray}
Now, having applied the formula for the double Borel transformation
\begin{eqnarray*}
&&\mathcal{B}_{M_{1}^{2}}\mathcal{B}_{M_{2}^{2}}\frac{(l-1)!}{\left[
m_{b}^{2}-(1-u)p^{2}-u(p+q)^{2}\right] ^{l}} \\
&=&\left( M^{2}\right) ^{2-l}e^{-m_{b}^{2}/M^{2}}\delta (u-u_{0}),
\end{eqnarray*}%
with
\begin{equation*}
u_{0}=\frac{M_{1}^{2}}{M_{1}^{2}+M_{2}^{2}},\ \ M^{2}=\frac{%
M_{1}^{2}M_{2}^{2}}{M_{1}^{2}+M_{2}^{2}},
\end{equation*}%
it is not difficult to find a desired expression for the Borel
transformation of the invariant amplitude in terms of the quark-gluon
degrees of freedom.

By this manner we obtain
\begin{eqnarray}
&&\mathcal{B}_{M_{1}^{2}}\mathcal{B}_{M_{2}^{2}}F^{QCD}(p^{2},%
\,(p+q)^{2})=e^{-m_{b}^{2}/M^{2}}  \notag \\
&&\times M^{2}\left\{ m_{b}F_{M}^{(s)}\phi _{M}^{(s)}(u_{0})\left( 1-\frac{%
m_{M}^{2}u_{0}\overline{u}_{0}}{M^{2}}\right) \right.   \notag \\
&&\left. +\frac{\phi _{3M}^{(s)p}(u_{0})}{2m_{s}}u_{0}+\frac{\phi
_{3M}^{(s)\sigma }(u_{0})}{6m_{s}}+\frac{1}{12m_{s}}u_{0}\frac{d\phi
_{3M}^{(s)\sigma }(u_{0})}{du}\right.   \notag \\
&&\left. +\frac{m_{b}^{2}\phi _{3M}^{(s)\sigma }(u_{0})}{6m_{s}M^{2}}+\frac{2%
{F_{M}^{(s)}}m_{b}}{M^{2}}u_{0}G_{4}(u_{0})-\frac{F_{M}^{(s)}m_{b}^{3}}{%
4M^{4}}\phi _{4M}^{(s)}(u_{0})\right.   \notag \\
&&\left. +2f_{3M}^{(s)}I_{M}^{3(s)}(u_{0})+F_{M}^{(s)}m_{b}\frac{%
I_{M}^{4(s)}(u_{0})}{M^{2}}\right\} ,  \label{eq:Bsum}
\end{eqnarray}%
In Eq.\ (\ref{eq:Bsum}) the new functions
\begin{eqnarray}
&&I_{M}^{3(s)}(u_{0})=\int_{0}^{u_{0}}d\alpha _{1}\left[ \frac{\Phi
_{3M}^{(s)}(\alpha _{1},1-u_{0},u_{0}-\alpha _{1})}{u_{0}-\alpha _{1}}%
\right.   \notag \\
&&\left. -\int_{u_{0}-\alpha _{1}}^{1-\alpha _{1}}d\alpha _{3}\frac{\Phi
_{3M}^{(s)}(\alpha _{1},1-\alpha _{1}-\alpha _{3},\alpha _{3})}{\alpha
_{3}^{2}}\right] ,
\end{eqnarray}%
and
\begin{eqnarray}
&&I_{M}^{4(s)}(u_{0})=\int_{0}^{u_{0}}d\alpha _{1}\int_{u_{0}-\alpha
_{1}}^{1-\alpha _{1}}\frac{d\alpha _{3}}{\alpha _{3}}\left[ 2\Psi
_{4M}^{(s)}(\alpha )-\Phi _{4M}^{(s)}(\alpha )\right.   \notag \\
&&\left. +2\widetilde{\Psi }_{4M}^{(s)}(\alpha )-\widetilde{\Phi }%
_{4M}^{(s)}(\alpha )\right]
\end{eqnarray}%
are introduced.

The Eq.\ (\ref{eq:Bsum}) is the required Borel transformed expression for the
function $F^{\mathrm{QCD}}(p^{2},(p+q)^{2})$ given in the quark-gluon
degrees of freedom. In order to derive the light-cone sum rule formulas for
the couplings $g_{B_{s}^{\ast }B_{s}\eta }$ and $g_{B_{s}^{\ast }B_{s}\eta
^{\prime }}$ one should equate Borel transformations of $F^{\mathrm{h}%
}(p^{2},(p+q)^{2})$ as in Eq.\ (\ref{eq:BHad}) and $F^{\mathrm{QCD}%
}(p^{2},(p+q)^{2})$ written down in Eq.\ (\ref{eq:Bsum}). Then the only
unknown term is a contribution of higher resonances and continuum states
represented in Eq.\ (\ref{eq:BHad}) as the integral with double spectral
density $\rho ^{\mathrm{h}}(s_{1},\,s_{2})$. To solve this problem, in
accordance with the main idea of the sum rule methods, we suggest that above
a some threshold in the $(s_1,\,s_2)$ plane the double spectral density $%
\rho ^{h}(s_{1},\,s_{2})$ can be replaced by $\rho ^{\mathrm{QCD}%
}(s_{1},\,s_{2})$. Then the continuum subtraction can be performed in accordance with
the procedure developed in Refs.\ \cite{Braun:1989,Braun:1995,Braun:1988qv}. It is based on the
observation that double spectral density in the leading contributions, i.e. in ones that are proportional to the positive powers of the Borel parameter $M^2$, is concentrated (or can be expanded) near the diagonal $s_1=s_2$. In this case for the continuum subtraction the simple expressions can be derived, which are not sensitive to the shape of the duality region \cite{Braun:1989,Braun:1995,Braun:1988qv}. The general formula in the case $M_{1}^{2}=M_{2}^{2}=2M^{2}$ and $u_{0}=1/2$ reads
\begin{equation}
M^{2n}e^{-\frac{m_b^{2}}{M^2}} \rightarrow \frac{1}{\Gamma(n)}\int_{m_b^{2}}^{s_0}
dse^{-\frac{s}{M^2}}\left ( s-m_b^{2}\right )^{n-1},\,n \geq 1.
\label{eq:CSub1}
\end{equation}
For terms  $\sim M^{2}$ it leads to the simple prescription
\begin{equation}
M^2e^{-m_{b}^{2}/M^{2}}\rightarrow M^2\left(
e^{-m_{b}^{2}/M^{2}}-e^{-s_{0}/M^{2}}\right),
\end{equation}
adopted in our work, as well.

For the higher-twist terms, which are proportional to zeroth or to the negative powers of $M^{2}$, on the one hand, continuum subtraction is not expected to have a large effect, and, on the other hand, it is not known how to perform it in theoretically clean way. The difficulty here is that the quark-hadron duality is not expected to work point-wise in the two-dimensional plane $(s_1,\,s_2)$, but, at best, after integration over the line $s_1+s_2=\rm {const}$ (see, for example Refs.\ \cite{Blok:1993,Ball:1994}). For this reason a naive subtraction using the "square" duality region $s_1<s_0,\,s_2<s_0$ does not have the strong theoretical basis. The spectral densities corresponding to the higher-twist terms under consideration are not concentrated near the diagonal $s_1=s_2$, as a result, the  required continuum subtractions take rather complicated forms. Because the higher-twist spectral densities decrease with $s_1$ and $s_2$ fast enough and an impact of the subtracted terms on the final result is not significant, in a standard technique of the LCSRs of this type one does not perform continuum subtractions in these terms at all \cite{Braun:1995}. Here we follow these procedures and subtract the continuum contributions only in the terms $\sim M^2$.

The masses of the $B_{s}$ and $B_{s}^{\ast}$ mesons are numerically close to each other, hence
in our calculations we can safely set $M_{1}^{2}=M_{2}^{2}$ and $u_0=1/2$. Then, it is not
difficult to write down the following sum rule:
\begin{eqnarray}
&&f_{B_{s}}f_{B_{s}^{\ast }}g_{B_{s}^{\ast }B_{s}M}=\frac{m_{b}}{%
m_{B_{s}}^{2}m_{B_{s}^{\ast }}}e^{\frac{m_{B_{s}}^{2}+m_{B_{s}^{\ast }}^{2}}{%
2M^{2}}}  \notag \\
&&\times \left\{ M^{2}\left( e^{-\frac{m_{b}^{2}}{M^{2}}}-e^{-\frac{s_{0}}{%
M^{2}}}\right) \left[ m_{b}F_{M}^{(s)}\phi _{M}^{(s)}(u_{0})\right. \right.
\notag \\
&&+\frac{\phi _{3M}^{(s)p}(u_{0})}{2m_{s}}u_{0}+\frac{\phi _{3M}^{(s)\sigma
}(u_{0})}{6m_{s}}  \notag \\
&&\left. +\frac{1}{12m_{s}}u_{0}\frac{d\phi _{3M}^{(s)\sigma }(u_{0})}{du}%
+2f_{3M}^{(s)}I_{M}^{3(s)}(u_{0})\right]   \notag \\
&&+e^{-\frac{m_{b}^{2}}{M^{2}}}\left[ F_{M}^{(s)}m_{b}\left( -m_{M}^{2}u_{0}%
\overline{u}_{0}\phi _{M}^{(s)}(u_{0})+2u_{0}G_{4M}^{(s)}(u_{0})\right.
\right.   \notag \\
&&\left. \left. \left. +I_{M}^{4(s)}(u_{0})-\frac{m_{b}^{2}}{4M^{2}}\phi
_{4M}^{(s)}(u_{0})\right) +\frac{m_{b}^{2}}{6m_{s}}\phi _{3M}^{(s)\sigma
}(u_{0})\right] \right\} _{u_{0}=1/2}.  \label{eq:SR}
\end{eqnarray}%
This result differs from the corresponding expression of Ref.\ \cite{Braun:1995} due to new definitions of the DAs, and the additional mass term in the sum rule expression.

For self-consistent treatment of Eq.\ (\ref{eq:SR}) one
needs expressions for  $f_{B_{s}}$ and $f_{B_{s}^{\ast }}$ with NLO accuracy. Recent calculation of the heavy-light mesons' decay constants, performed in the context of QCD sum rules method by taking into account $O(\alpha_s^{2})$ terms in the perturbative part and  $O(\alpha_s)$ corrections to the quark-condensate contribution, can be found in   Ref.\ \cite{Gelhausen:2013wia}. For further details and explicit expressions we refer to this work (see, also \cite{Wang:2015mxa}).


\subsection{NLO corrections. Gluonic contributions to the strong couplings}

The QCD LCSR for the strong couplings Eq.\ (\ref{eq:SR}) have been derived at the
leading order of the perturbative QCD with twist-4 accuracy. In order to
improve our results and make more precise theoretical predictions for the
strong couplings we need to find NLO perturbative corrections at least to
the leading twist term, and by this way include into analysis also the gluon
component of the eta mesons.
\begin{figure}[t]
\centerline{
\begin{picture}(210,140)(0,0)
\put(-5,20){\epsfxsize7.8cm\epsffile{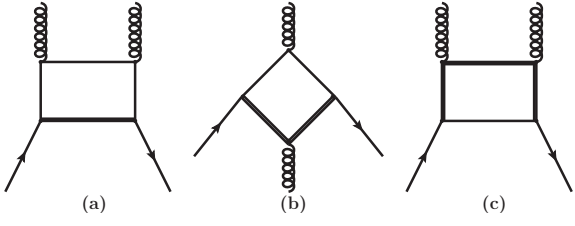}}
\end{picture}
}
\caption{ Quark-box diagrams that determine the gluonic contribution. Thick
lines correspond to a heavy quark.}
\label{fig:LCSR-2}
\end{figure}
The NLO correction to the leading twist term, and relevant double spectral
density for the strong vertices $B^{\ast }B\pi $ and $D^{\ast
}D\pi $ were found in Ref.\ \cite{Khodjamirian:1999}. In this work authors
demonstrated that, to this end, it is sufficient to utilize NLO correction
to the transition form factor $B\rightarrow \pi $ calculated in Ref.\ \cite%
{Khodjamirian:1997}, and from the corresponding expression deduced the
double spectral density for the coupling $g_{B^{\ast }B\pi }$. Because the
pion is a pseudoscalar particle, and has only quark component, after some
corrections that depend on definitions of  DAs and decay constants,
results of this work can be used to find NLO corrections to the leading twist
term in the LCSRs for strong couplings arising from the quark component of
the $\eta $ and $\eta ^{\prime }$ mesons. Therefore, we borrow corresponding
expression for the NLO correction from the work \cite{Khodjamirian:1999},
and for the asymptotic DAs  $\phi^{(s)}_{\eta(\eta^{\prime})}(u)$
 get:
\begin{eqnarray}
&&Q_{\eta (\eta ^{\prime })}^{(s)}\left( M^{2},\ s_{0}^{B_{s}}\right) =\frac{%
\alpha _{s}C_{F}}{4\pi }\frac{F_{\eta (\eta ^{\prime })}^{(s)}m_{b}}{\sqrt{2}%
}  \notag \\
&&\times \int_{2m_{b}^{2}}^{2s_{0}^{B_{s}}}f\left( \frac{s}{m_{b}^{2}}%
-2\right) e^{-s/2M^{2}}ds,  \label{eq:SRQ}
\end{eqnarray}%
where%
\begin{eqnarray}
&&f(x)=\frac{\pi ^{2}}{4}+3\ln \left( \frac{x}{2}\right) \ln \left( 1+\frac{x%
}{2}\right)  \nonumber \\
&&-\frac{3\left( 3x^{3}+22x^{2}+40x+24\right) }{3(2+x)^{3}}\ln \left( \frac{x%
}{2}\right) \nonumber  \\
&&+6\mathrm{Li}_{2}\left( -\frac{x}{2}\right) -3\mathrm{Li}_{2}(-x)-3\mathrm{%
Li}_{2}(-1-x) \nonumber \\
&&-3\ln (1+x)\ln (2+x)-\frac{3(3x^{2}+20x+20)}{4(2+x)^{3}} \\
&&+\frac{6x(1+x)\ln (1+x)}{(2+x)^{2}}.
\end{eqnarray}

In order to find the gluonic contributions to the LCSRs one has to compute the
quark-box diagrams shown in Fig.\ \ref{fig:LCSR-2}. For the transitions $%
B\rightarrow \eta ^{(\prime )}$ they were calculated in Ref.\ \cite%
{Ball:2007hb} (see also, \cite{Dupl:2015}). We adapt to our problem the
relevant expressions  obtained in Ref.\ \cite%
{Ball:2007hb} and use them in our calculations.

To derive the double spectral density, we start from the expression
\begin{equation}
F^{(g)}( p^{2},\, (p+q)^{2} )=\frac{\alpha _{s}C_{F}}{4\pi }%
f_{M}^{(1)}m_{b}\int_{m_{b}^{2}}^{\infty }\frac{d\alpha g(\alpha ,p^{2})}{%
\alpha -(p+q)^{2}},
\end{equation}%
where
\begin{eqnarray}
&&g(\alpha ,p^{2})=\frac{25}{6\sqrt{3}}a_{2,M}^{(g)}\left\{ \frac{%
m_{b}^{2}-\alpha }{(\alpha -p^{2})^{5}}\left[ 59m_{b}^{6}\right. \right. \nonumber \\
&&+21p^{6}-63p^{4}\alpha -19p^{2}\alpha ^{2}+2\alpha ^{3}+m_{b}^{2}\alpha
\left( 164p^{2}+13\alpha \right)  \nonumber \\
&&\left. -m_{b}^{4}\left( 82p^{2}+95\alpha \right) \right] +6\frac{%
(m_{b}^{2}-p^{2})(\alpha -m_{b}^{2})}{(\alpha -p^{2})^{5}} \nonumber \\
&&\times \left[ 5m_{b}^{4}+p^{4}+3p^{2}\alpha +\alpha
^{2}-5m_{b}^{2}(p^{2}+\alpha )\right]  \nonumber \\
&&\left. \times \left[ 2\ln \frac{\alpha -m_{b}^{2}}{m_{b}^{2}}-\ln \frac{%
\mu ^{2}}{m_{b}^{2}}\right] \right\} .
\end{eqnarray}%
We employ a method described in detailed form in Ref.\ \cite{Ball:1994}. In
other words, first we perform the double Borel transformations
\begin{eqnarray*}
&&B_{t_{1}}(p^{2})B_{t_{2}}((p+q)^{2})F^{(g)}(p^{2},\ (p+q)^{2} )\equiv \hat{F}%
^{(g)}(t_{1},t_{2}) \\
&=&\frac{1}{t_{1}t_{2}}\int ds_{1}ds_{2}\rho
(s_{1},s_{2})e^{-s_{1}/t_{1}-s_{1}/t_{2}},
\end{eqnarray*}%
then apply the Borel transformations in $\tau _{1}=1/t_{1}$ and $\tau
_{2}=1/t_{2}$ in order to extract $\rho (s_{1},s_{2})$
\begin{equation*}
B_{1/s_{1}}(\tau _{1})B_{1/s_{2}}(\tau _{2})\frac{1}{\tau _{1}\tau _{2}}\hat{%
F}^{(g)}(1/\tau _{1},1/\tau _{2})=s_{1}s_{2}\rho(s_{1},s_{2}).
\end{equation*}%
Having subtracted contribution of the resonances and continuum states we get the
gluonic correction as the double dispersion integral:
\begin{eqnarray}
&&F_{M}\left(  p^{2},\ (p+q)^{2} \right)=\frac{\alpha _{s}C_{F}}{%
4\pi }f_{M}^{(1)}m_{b}  \notag \\
&&\times \int_{m_{b}^{2}}^{s_{0}^{B_{s}}}\int_{m_{b}^{2}}^{s_{0}^{B_{s}}}%
\frac{ds_{1}ds_{2}\rho (s_{1},s_{2})}{(s_{1}-p^{2})(s_{2}-(p+q)^{2})},
\label{eq:SRG}
\end{eqnarray}%
where
\begin{equation*}
\rho (s_{1},s_{2})=\frac{25}{6\sqrt{3}}a_{2,M}^{(g)}\left[ \rho
_{1}(s_{1},s_{2})+6\rho _{2}(s_{1},s_{2})\right] .
\end{equation*}%
Here
\begin{eqnarray}
&&\rho _{1}(s_{1},s_{2})=21\Delta ^{(1)}(s_{1}-s_{2})  \notag \\
&&-\frac{82}{6}\Delta ^{{(3)}}(s_{1}-s_{2})-\frac{59}{24}\Delta
^{(4)}(s_{1}-s_{2}),  \label{eq:spect1}
\end{eqnarray}%
and
\begin{eqnarray}
&&\rho _{2}(s_{1},s_{2})=L(s_{1},\mu )\left[ \Delta
^{(2)}(s_{1}-s_{2})\right.   \notag \\
&&\left. +\frac{1}{3}\Delta ^{(3)}(s_{1}-s_{2})+\frac{1}{24}\Delta
^{(4)}(s_{1}-s_{2})\right] .  \label{eq:spect2}
\end{eqnarray}%
In Eqs.\ (\ref{eq:spect1}) and (\ref{eq:spect2})
\begin{eqnarray}
\Delta ^{(n)}(s_{1}-s_{2}) &=&(s_{1}-m_{b})^{n}\delta ^{(n)}(s_{1}-s_{2}),
\notag \\
L(s,\mu ) &=&2\ln \frac{s-m_{b}^{2}}{m_{b}^{2}}-\ln \frac{\mu ^{2}}{m_{b}^{2}%
},
\end{eqnarray}%
with $\delta ^{(n)}(s_{1}-s_{2})$ being defined as
\begin{equation*}
\delta ^{(n)}(s_{1}-s_{2})=\frac{\partial ^{n}}{\partial s_{1}^{n}}\delta
(s_{1}-s_{2}).
\end{equation*}%
The Borel transformations in the variables $p^{2}$ and $(p+q)^{2}$ of the
integral in Eq.\ (\ref{eq:SRG}) gives us the desired gluonic contribution to the
sum rules
\begin{eqnarray}
&&\mathcal{B}_{M_{1}^{2}}\mathcal{B}_{M_{2}^{2}}F_{M}\left( (p+q)^{2},\ p^{2} \right)  =\frac{\alpha _{s}C_{F}}{4\pi }f_{M}^{(1)}m_{b} \nonumber \\
&&\times \int_{m_{b}^{2}}^{s_{0}^{B_{s}}}ds_{1}%
\int_{m_{b}^{2}}^{s_{0}^{B_{s}}}ds_{2}\rho
(s_{1},s_{2})e^{-s_{1}/M_{1}^{2}}e^{-s_{2}/M_{2}^{2}}.
\label{eq:GBor}
\end{eqnarray}%
In the case $M_{1}^{2}=M_{2}^{2}=2M^{2}$ by applying methods from
Appendix B of Ref.\ \cite{Braun:1995} , we calculate the integrals in Eq.\ (\ref{eq:GBor})
\begin{eqnarray}
&&\int_{m_{b}^{2}}^{s_{0}^{B_{s}}}ds_{1}%
\int_{m_{b}^{2}}^{s_{0}^{B_{s}}}ds_{2}\Delta
^{(k)}(s_{1}-s_{2})e^{-(s_{1}+s_{2})/2M^{2}} \nonumber \\
&&=\frac{(-1)^{k}}{2^{k+1}}\int_{2m_{b}^{2}}^{2s_{0}^{B_{s}}}dse^{-s/2M^{2}}%
\left( \frac{d}{dv}\right) ^{k}\left( v-\frac{m_{b}^{2}}{s}\right)
_{v=1/2}^{k}
\end{eqnarray}%
and
\begin{eqnarray}
&&\int_{m_{b}^{2}}^{s_{0}^{B_{s}}}ds_{1}%
\int_{m_{b}^{2}}^{s_{0}^{B_{s}}}ds_{2}\ln \left( s_{1}-m_{b}^{2}\right)
\Delta ^{(k)}(s_{1}-s_{2}) \nonumber \nonumber \\
&&\times e^{-(s_{1}+s_{2})/2M^{2}}=\frac{(-1)^{k}}{2^{k+1}}\int_{2m_{b}^{2}}^{2s_{0}^{B_{s}}}dse^{-s/2M^{2}}%
\left( \frac{d}{dv}\right) ^{k}  \nonumber \nonumber \\
&&\times \left[ \left( v-\frac{m_{b}^{2}}{s}\right)
^{k}\ln \left( sv-m_{b}^{2}\right) \right] _{v=1/2}.
\end{eqnarray}%
The integrations over $s$ can be performed explicitly that
allows us to find the gluonic contribution in a rather simple form
\begin{eqnarray}
&&\widetilde{Q}_{\eta (\eta ^{\prime })}\left( M^{2},\ s_{0}^{B_{s}}\right)
=\frac{\alpha _{s}C_{F}}{4\pi }f_{\eta (\eta ^{\prime })}^{(1)}m_{b} \nonumber \\
&&\times \left[ r_{1}(M^{2},s_{0}^{B_{s}})+r_{2}(M^{2},s_{0}^{B_{s}})\right] ,
\end{eqnarray}%
where%
\begin{equation}
r_{1}(M^{2},s_{0}^{B_{s}})=M^{2}\left(
e^{-m_{b}^{2}/M^{2}}-e^{-s_{0}/M^{2}}\right) \left( -\frac{51}{32}\right) ,
\end{equation}%
and%
\begin{eqnarray}
&&r_{2}(M^{2},s_{0}^{B_{s}}) =\frac{3}{16}M^{2}e^{-m_{b}^{2}/M^{2}}\left[
22+20\psi (7)
\right.  \nonumber \\
&&\left. -20\Gamma \left( 0,\frac{s_{0}^{B_{s}}-m_{b}^{2}}{M^{2}}\right)+20\ln \frac{2M^{2}}{m_{b}^{2}}-10\ln \frac{\mu ^{2}}{m_{b}^{2}}%
\right]  \nonumber \\
&&+\frac{3}{16}M^{2}e^{-s_{0}^{B_{s}}/M^{2}}\left[ -27-20\ln \frac{%
2\left( s_{0}^{B_{s}}-m_{b}^{2}\right) }{m_{b}^{2}} \right. \nonumber \\
&& \left. +10\ln \frac{\mu^{2}}{m_{b}^{2}}\right] .
\end{eqnarray}%
Here $\psi (z)=(d/dz)\ln \Gamma (z)$ and  $\Gamma (a,z)$ are digamma and
incomplete Gamma functions, respectively.

Then the NLO corrections to LCSRs arising from the quark and gluonic
components of the eta mesons are given by the expression%
\begin{equation}
\frac{m_{b}}{m_{B_{s}}^{2}m_{B_{s}^{\ast }}}e^{\frac{%
m_{B_{s}}^{2}+m_{B_{s}^{\ast }}^{2}}{2M^{2}}}\left( Q_{\eta (\eta ^{\prime
})}^{(s)}+\widetilde{Q}_{\eta (\eta ^{\prime })}\right),
\label{eq:NLOSR}
\end{equation}%
which should be added to Eq.\ (\ref{eq:SR}).

\begin{figure}[t]
\centerline{
\begin{picture}(210,140)(0,0)
\put(-5,0){\epsfxsize7.8cm\epsffile{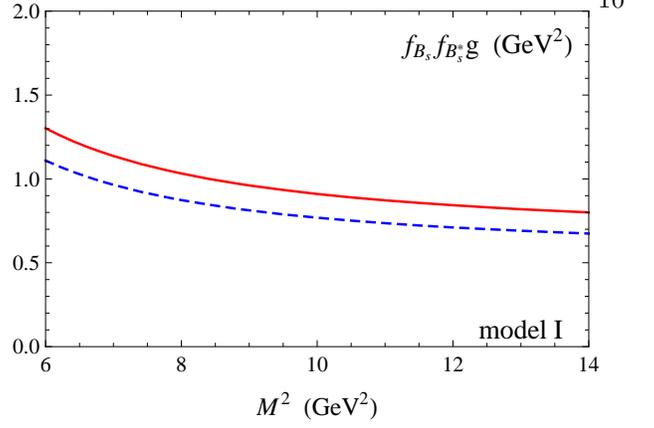}}
\end{picture}}
\caption{The strong couplings as functions of the Borel parameter $M^2$. The solid (red) line describes $f_{B_{s}}f_{B_{s}^{\ast}}g_{B_{s}^{\ast}B_{s}\eta^{\prime}}$, whereas the dashed (blue) curve corresponds to $ f_{B_{s}}f_{B_{s}^{\ast}}\mid g_{B_{s}^{\ast}B_{s}\eta}\mid$. In computations the model ${\rm I}$ is used. The parameter $s_0^{B_{s}}$ is set equal to $36\ {\rm GeV}^2$.}
\label{fig:G}
\end{figure}
It is interesting to note that strong couplings given by Eqs.\ (\ref{eq:SR}) and (\ref{eq:NLOSR}) may be presented in the form
\begin{eqnarray}
&&g_{B_{s}^{\ast}B_{s}\eta }\simeq -\sin \varphi \ G_{B_{s}^{\ast}B_{s}
\eta }^{(s)},\,  \notag \\
&&g_{B_{s}^{\ast}B_{s}\eta ^{\prime }}\simeq \cos \varphi \
G_{B_{s}^{\ast}B_{s}\eta ^{\prime }}^{(s)}.
\end{eqnarray}%
In fact,  excluding some terms, the couplings with the high accuracy follow the mixing pattern discussed above that can be demonstrated explicitly.


\section{Numerical results and conclusions}

\label{sec:results}
\begin{figure}
\centerline{
\begin{picture}(210,170)(0,0)
\put(-5,5){\epsfxsize7.8cm\epsffile{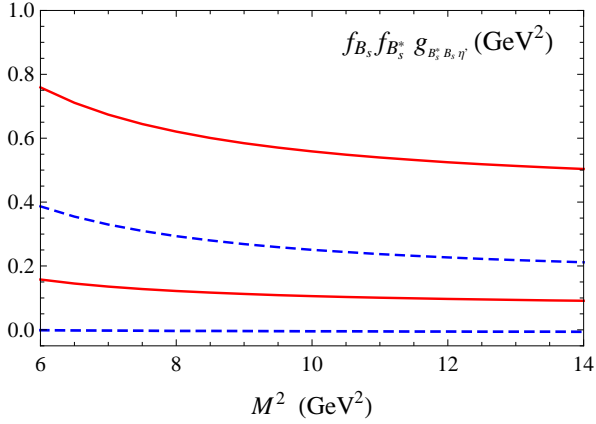}}
\end{picture}
}
\caption{Contributions to the coupling $f_{B_{s}}f_{B_{s}^{\ast}}g_{B_{s}^{\ast}B_{s}\eta^{\prime}}$ originating from the leading, the higher-twist and NLO terms. The upper solid (red) line is contribution of LO twist-2 term, the upper dashed line (blue) shows contribution of the higher-twist terms, the lower solid (red) curve is the NLO effect coming from the meson's quark component, and the lower dashed (blue) line is the gluonic contribution to the coupling. The parameters are the same as in Fig.\ \ref{fig:G}.}
\label{fig:OrHT}
\end{figure}

The LCSR expressions for $g_{B^{*}_sB_{s}\eta}$ and $g_{B^{*}_sB_{s}\eta^{\prime}}$
in Eqs.\ (\ref{eq:SR}) and (\ref{eq:NLOSR})
contain numerous parameters that should be
fixed in accordance with the usual procedures. But apart from that in numerical
calculations there is a necessity to utilize also equalities to connect $\eta
$ and $\eta^{\prime}$ mesons' DAs and decay constants obtained using
different bases. Indeed, as we have emphasized above, in order to solve
renormalization group equations it is convenient to use the singlet-octet
basis. This basis was used in Ref.\ \cite{Agaev:2014} to describe evolution
of the flavor-octet and flavor-singlet DAs with NLO accuracy. One should
note that the gluon DA in Eq.\ (\ref{gluonDA}) is normalized in terms of the
decay constant $f_{M}^{(1)}$. From another side, the QF basis is more
suitable to analyze the $\eta-\eta^{\prime}$ mixing phenomena and solve
equations of motions, which determine parameters in twist-4 DAs. The values
of the decay constants in Eq.\ (\ref{FKSvalues}) were deduced within the QF
mixing scheme, as well. The general expression for such transformations can
be found in Eq.\ (\ref{relation}). Here we provide the formula for eta
mesons' decay constants in the SO basis
\begin{align*}
\begin{pmatrix}
f_\eta^{(8)} & f_\eta^{(1)} \\
f_{\eta^{\prime}}^{(8)} & f_{\eta^{\prime}}^{(1)}%
\end{pmatrix}
=&
\begin{pmatrix}
\cos\theta_8 & -\sin\theta_1 \\
\sin\theta_8 & \cos\theta_1%
\end{pmatrix}
\begin{pmatrix}
f_8 & 0 \\
0 & f_1%
\end{pmatrix}
\label{relation}
\end{align*}
with the numerical values of the parameters
\begin{eqnarray*}
&&f_1=(1.17\pm 0.03)f_{\pi},\,\,f_8=(1.26\pm 0.04)f_{\pi}, \\
&& \theta_1=-(9.2^{\circ} \pm 1.7^{\circ}),\,\, \theta_8=-(21.2^{\circ} \pm
1.6^{\circ}).
\end{eqnarray*}
The $B_s$ and $B_{s}^{\ast}$ mesons' decay constants and masses enter to
Eqs.\ (\ref{eq:SR}) and (\ref{eq:NLOSR}) as input parameters.
 Their values are collected below (in $\mathrm{MeV}$)
\begin{eqnarray*}
&&m_{\eta}=547.86 \pm 0.02,\,\,\, m_{\eta^{\prime}}=957.78 \pm 0.06, \\
&&m_{B_{s}}=5366.77 \pm 0.4, \,\, \, m_{B_{s}^{\ast}}=5415.4 \pm 1.5.
\end{eqnarray*}
The decay constants $f_{B_{s}}$ and $f_{B_{s}^{\ast}}$ were calculated from
the two-point QCD sum rules in Ref.\ \cite{Wang:2015mxa} (in $\mathrm{MeV}$)
\begin{equation}
f_{B_{s}}=231 \pm 16, \,\, f_{B_{s}^{\ast}}=213 \pm 18.
\end{equation}
We employ masses of the quarks in the $\overline{MS}$ scheme (in \textrm{GeV})
\begin{equation}
m_b(m_b)=4.18 \pm 0.03,\, m_c(m_c)=1.275 \pm 0.025,
\end{equation}
Their scale dependencies are taken into account
in accordance with the renormalization group evolution
\begin{equation*}
m_{q}(\mu)=m_{q}(\mu_0)\left [ \frac{\alpha_s(\mu)}{\alpha_s(\mu_0)}\right
]^{\gamma_{q}},
\end{equation*}
with $\gamma_b=12/23$ and $ \gamma_c=12/25$.
The strange quark mass is $m_s=0.137\ \, {\rm GeV}$.
The renormalization scale is set equal to
\begin{equation}
\mu_b=\sqrt {m_{B_{s}}^2-m_{b}^{2}}\simeq 3.4\,\,\mathrm{GeV}.
\end{equation}
The parameters and quantities are evolved to this scale employing the
two-loop QCD running coupling $\alpha_{s}(\mu)$ with $\Lambda^{(4)}=326\,\,%
\mathrm{MeV}$. The same QCD two-loop coupling is used throughout this work,
for example, to compute NLO corrections. The evolution of the leading twist
DAs is calculated with the NLO accuracy by taking into account quark-gluon mixing
\cite{Agaev:2014}.
\begin{figure}
\centerline{
\begin{picture}(210,150)(0,0)
\put(-5,5){\epsfxsize7.8cm\epsffile{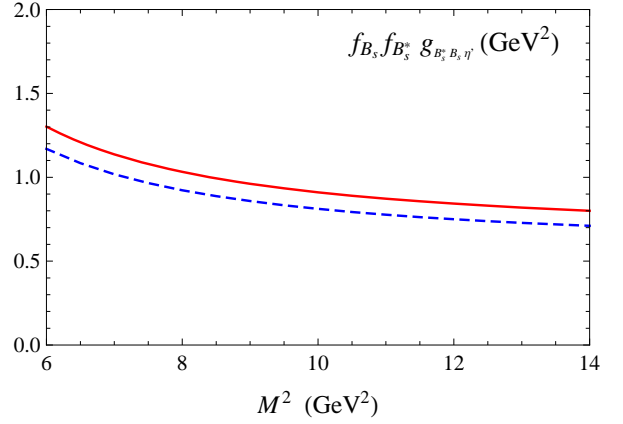}}
\end{picture}
}
\caption{The coupling $f_{B_{s}}f_{B_{s}^{\ast}}g_{B_{s}^{\ast}B_{s}\eta^{\prime}}$ computed using the different model DAs. Correspondence between the curves and models is: the solid (red) line\ -\ model ${\rm I}$  and the dashed (blue) line\ - \ model ${\rm III}$.}
\label{fig:DifDA}
\end{figure}
\begin{figure}[t]
\centerline{
\begin{picture}(210,170)(0,0)
\put(-5,5){\epsfxsize7.8cm\epsffile{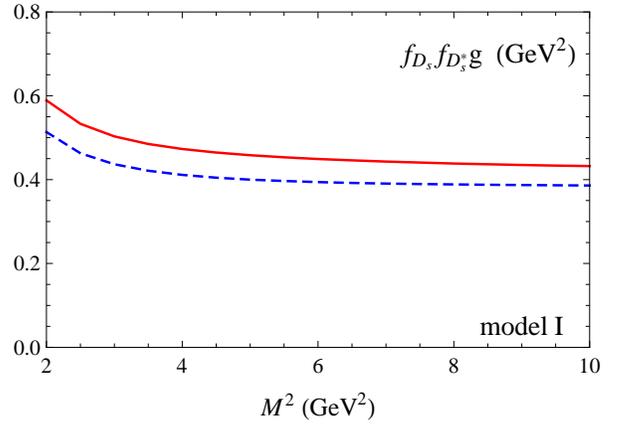}}
\end{picture}
}
\caption{The couplings as functions of the Borel parameter $M^2$. The solid (red) line corresponds to $f_{D_{s}}f_{D_{s}^{\ast}}g_{D_{s}^{\ast}D_{s}\eta^{\prime}}$, the dashed (blue) curve is the coupling
$f_{D_{s}}f_{D_{s}^{\ast}}\mid g_{D_{s}^{\ast}D_{s}\eta}\mid$. In computations the model ${\rm I}$ is used.
The parameter $s_0^{D_{s}}$ is set equal to $7\ {\rm GeV}^2$.}
\label{fig:Charm}
\end{figure}
Calculations require to fix the
threshold parameter $s_{0}$ and a region within of which it may be varied.
For $s_{0}$ we employ
\begin{equation*}
s_{0}^{B_{s}}\equiv s_{0}^{B_{s}^{\ast}}\simeq 36 \pm 2.5\,\,{\rm GeV}^2.
\end{equation*}
Additionally, the eta mesons' DAs contain the Gegenbauer moments
$a_{n}^{(1,8)}(\mu_0)$ and $a_{n}^{(g)}(\mu_0)$. In Ref.\ \cite{Agaev:2014} they were
extracted from the analysis of the eta mesons' electromagnetic transition form factors. In the present
work for $a_{n}^{(1,8)}$ and $a_{2}^{(g)}$ we utilize values that are compatible with ones
from this work and accept the following models for DAs
\begin{eqnarray}
&&{\rm I}. \,\,\,a_{2}^{(1,8)}=a_{4}^{(1,8)}=0.1,\,\,a_{2}^{(g)}=- 0.2,\nonumber \\
&& {\rm II}.\,\,\,a_{2}^{(1,8)}=a_{4}^{(1,8)}=0.2,\,\,a_{2}^{(g)}=- 0.2, \nonumber \\
&& {\rm III}. \,\,\,a_{2}^{(1,8)}=0.2,\,a_{4}^{(1,8)}=0,\,\,a_{2}^{(g)}=- 0.2.
\label{eq:Mod1}
\end{eqnarray}

Results of the computations of the "scaled" couplings $f_{B_{s}}f_{B_{s}^{\ast}}g_{B_{s}^{\ast}B_{s}\eta^{\prime}}$ and $ f_{B_{s}}f_{B_{s}^{\ast}}\mid g_{B_{s}^{\ast}B_{s}\eta}\mid$ are depicted in Fig. \ref{fig:G}. Calculations have been carried out employing the model ${\rm I}$.  From analysis we find the range of values of the Borel parameter $8\ {\rm GeV}^2<M^2<12\ {\rm GeV}^2$, where the effects of the higher resonances and continuum states is less than  $30\%$ of the leading order twist-2 contribution, and  terms $\sim M^{-2}$ form only $\sim 5\%$ of the sum rule. Additionally, in this interval the dependence of the couplings on $M^2$ is stable, and one may expect that the sum rule gives the reliable predictions.

The sum rules  receive contributions from the different terms that are shown in Fig. \ref{fig:OrHT}.  The main component is the leading order twist-2 term: it forms approximately $ 60 \%$ of the strong couplings. The effect of the NLO quark correction is also essential: in the explored range of the Borel parameter it equals to $\simeq 12.5 \%$ of the coupling $f_{B_{s}}f_{B_{s}^{\ast}}g_{B_{s}^{\ast}B_{s}\eta^{\prime}}$. The same estimation is valid for  $f_{B_{s}}f_{B_{s}^{\ast}}g_{B_{s}^{\ast}B_{s}\eta}$, as well. Correction originating from the gluon content of the meson is very small. In fact, it equals  only to $\simeq -0.5\ \%$ of $f_{B_{s}}f_{B_{s}^{\ast}}g_{B_{s}^{\ast}B_{s}\eta^{\prime}}$.

The higher-twist terms play an essential role in forming of the couplings. Indeed, $\sim 28\% $ of their values within considering range of $M^2$ are due to HT corrections. The main part of the HT corrections are determined by the two-particle twist-3 DAs $\phi_{3\eta^{\prime}}^{(s)p}(u)$ and  $\phi_{3\eta^{\prime}}^{(s)\sigma}(u)$: they give $\sim 33\ \%$, whereas corrections of remaining HT terms are small $-5\%$.

The extracted  couplings, in general, depend on the distribution amplitudes utilized in calculations. We have computed the couplings using the different model DAs, and drown the results in Fig. \ref{fig:DifDA}. Some of the DAs  (models ${\rm I}$ and ${\rm II}$) lead to almost identical predictions such that  corresponding lines  become undistinguishable. Therefore, in Fig. \ref{fig:DifDA} we show only the line corresponding to the model ${\rm I}$.  At the same time, the results for couplings due to  another pair of DAs (models ${\rm I}$ and ${\rm III}$) differ from each other considerably .

The predictions in the present work are made employing the model ${\rm I}$. By varying the parameters within the allowed ranges we estimate uncertainties of computations. The important sources of uncertainties are $M^2$ and $s_{0}^{B_{s}}$, as well as the decay constants $f_{B_{s}}$ and $f_{B_{s}^{\ast}}$ calculated within the two-point QCD sum rules. Having changed $M^2$ and $s_{0}^{B_{s}}$ within $8\ <M^2<12\ {\rm GeV}^2$, and $33.\ 5 <s_{0}^{B_{s}}\ <38.\ 5\ \,{\rm GeV}^2$ respectively, and taken into account uncertainties arising from the meson decay constants we get
\begin{eqnarray}
&&f_{B_{s}}f_{B_{s}^{\ast}}\mid g_{B_{s}^{\ast}B_{s}\eta}\mid=0.837 \pm 0.08\  {\rm GeV}^2,  \nonumber \\
&&f_{B_{s}}f_{B_{s}^{\ast}}g_{B_{s}^{\ast}B_{s}\eta^{\prime}}=0.994 \pm 0.12\ {\rm GeV}^2.
\end{eqnarray}
Dividing the product of the couplings by the decay constants gives for the couplings the following predictions:
\begin{equation}
\mid g_{B_{s}^{\ast}B_{s}\eta}\mid=17.08 \pm 1.63, \, \,g_{B_{s}B_{s}^{\ast}\eta^{\prime}}=20.2 \pm 2.44.
\end{equation}

We proceed in our studies and extract the strong couplings $g_{D^{*}_sD_{s}\eta}$ and $%
g_{D^{*}_sD_{s}\eta^{\prime}}$ (see, Fig.\ \ref{fig:Charm}). To this end, in all expressions we have to replace
$b \to c$.
The masses and decay constants in units of $\mathrm{MeV}$ are:
\begin{eqnarray}
&&m_{D_{s}}=1969 \pm 1.4, \,\, \, m_{D_{s}^{\ast}}=2112.1 \pm 0.4,  \notag \\
&&f_{D_{s}}=240 \pm 10, \,\, f_{D_{s}^{\ast}}=308 \pm 21.
\end{eqnarray}
All parameters should be adjusted to the new problem. This leads
to the replacements
\begin{equation}
\mu_c=\sqrt {m_{D_{s}}^2-m_{c}^{2}}\simeq 1.68\, \,\,{\rm GeV},
\end{equation}
and $s_{0}^{D_{s}}=7\pm 1\,\,\mathrm{GeV}^2$.
It has been found that the range of the Borel parameter $3\ {\rm GeV}^2<M^2<5\ {\rm GeV}^2$ is suitable for evaluating the sum rules. From the relevant sum rules for the product of the decay constants and coupling we extract the following values
\begin{eqnarray}
&&f_{D_{s}}f_{D_{s}^{\ast}}\mid g_{D_{s}^{\ast}D_{s}\eta}\mid=0.411 \pm 0.04\  {\rm GeV}^2,  \nonumber \\
&&f_{D_{s}}f_{D_{s}^{\ast}}g_{D_{s}^{\ast}D_{s}\eta^{\prime}}=0.473 \pm 0.042\ {\rm GeV}^2.
\end{eqnarray}
Then for the couplings we get
\begin{equation}
\mid g_{D_{s}^{\ast}D_{s}\eta}\mid=4.51 \pm 0.44, \, \,g_{D_{s}^{\ast}D_{s}\eta^{\prime}}=5.19 \pm 0.46.
\end{equation}

Our results have been obtained within the quark-hadron duality ansatz of
\cite{Braun:1995}, where $g_{D^{*}D\pi}$ and $g_{B^{*}B\pi}$ were evaluated. But there is a discrepancy between the predictions for $g_{D^{*}D\pi}$ and data of CLEO Collaboration \cite{CLEO:2002}. One of the main input parameters in these calculations is a value of the leading twist DA at $u_0=1/2$.  In Ref.\ \cite{Braun:1995} it was chosen as $\phi_{\pi}(1/2)\simeq 1.2$,  whereas recent analysis of the pion electromagnetic transition form factor performed in Refs.\ \cite{Agaev:2010aq,Agaev:2012tm} predicts LT pion DAs enhanced at the middle point: these model DAs at $u_0=1/2$ are very close to the asymptotic DA with $\phi_{\rm{asy}}(1/2)=1.5$. The usage of updated twist-3 DAs may also lead to sizeable corrections, because twist-3 terms contribute to  $g_{D^{*}D\pi}$ at the level of $(50-60) \%$, and are as important as the twist-2 term. All these questions necessitate new, updated investigation of the couplings $g_{D^{*}D\pi}$ and $g_{B^{*}B\pi}$ in the context of LCSRs method. The real accuracy of this method is not completely clear at present. On the one hand, it leads to results with $30-50 \%$ deviation from experimental data as in $g_{D^{*}D\pi}$ case, on the other hand, gives rather precise predictions for radiative decays of mesons. Indeed, LCSR prediction for $g_{D^{*}D\gamma}$  \cite{Aliev:1995wi,Rohrwild:2007yt} correctly describe experimental data: the value of the quark condensate' magnetic susceptibility that enters to this sum rule as a nonperturbative parameter is known from both QCD sum rules and lattice computations  \cite{Bali:2012jv} and agree with each other. As QCD lattice simulations of $g_{D^{*}D\pi}$ (see, Ref.\ \cite{Damir:2013}) agree with the CLEO data, it will be instructive to compare our predictions for the strong couplings  $g_{B^{*}_sB_{s}\eta^{(\prime)}}$ and $g_{D^{*}_sD_{s}\eta^{(\prime)}}$ with relevant lattice results, when they will be available.

The couplings $g_{B_{s}^{\ast}B_{s}\eta^{(\prime)}}$ were calculated in Ref.\ \cite{Yazici:2013eia} by applying the three-point sum rule method, as well. Differences in adopted definitions for
the couplings, chosen structures and  explored kinematical regimes to extract their values  make direct comparison of relevant findings rather problematic: we note only a sizeable numerical discrepancy between our predictions
and results of Ref.\ \cite{Yazici:2013eia}.
We emphasize also the advantage of the LCSR method compared to the three-point sum rules approach in calculations of the strong couplings or/and form factors. Indeed, in the three-point sum rules the
 higher orders in the operator product expansion (OPE) are enhanced by powers of the heavy quark mass and for sufficiently large masses the OPE breaks down. The LCSR method does not suffer from such problems: It is consistent with heavy-quark limit,
and provides more elaborated tools for investigations, than alternative approaches.

In the present work we have investigated the strong $D^{*}_sD_{s}\eta^{(\prime)}$ and  $%
B^{*}_sB_{s}\eta^{(\prime)}$ vertices and calculated the relevant couplings using the method of QCD sum rules on the light-cone. We have included into our analysis effects of the eta mesons' gluon components. The derived expressions has been explored and numerical values of the strong couplings $g_{D_{s}^{\ast}D_{s}\eta^{(\prime})}$ and $g_{B_{s}^{\ast}B_{s}\eta^{(\prime)}}$ have been evaluated. Studies have demonstrated that the direct contribution to the strong couplings arising from the two-gluon components of the $\eta$ and $\eta^{\prime}$ is small. But owing to mixing the gluon components affect the quark DAs, which  can not be ignored.

\section*{ACKNOWLEDGEMENTS}

S.~S~A. is grateful to T.~M.~ Aliev and V.~M.~Braun for enlightening and helpful
discussions. S.~S.~A. also thanks colleagues from the Physics Department of
Kocaeli University for warm hospitality. The work of S.~S.~A. was supported by the
Scientific and Technological Research Council of Turkey (TUBITAK) grant 2221-"Fellowship Program For Visiting Scientists and Scientists on
Sabbatical Leave".

\appendix*
\section{A}
\renewcommand{\theequation}{\Alph{section}.\arabic{equation}}

\label{sec:App}
This appendix is devoted to calculation of $f_{3s}$ and $\delta_{M}^{2(s)}$,
which enter as parameters into higher twist DAs of the $\eta$ and $%
\eta^{\prime}$ mesons. To this end, in the  two-point sum rules written down below,
we consider $f_{s}$ and $h_{s}$, as well as mixing angle $\varphi $ as input parameters; then only $f_{3s}
$ and $\delta_{M}^{2(s)}$ remain unknown.

The $f_{3s}$ and $\delta_{M}^{2(s)}$ can be defined in terms of matrix
elements of some local operators. Indeed, the parameter $f_{3s}$ can be
defined through the matrix element of the following twist-3 operator
\begin{equation*}
\langle 0\mid \overline{s}\sigma _{z\nu }\gamma _{5}gG_{z\nu }s\mid
M(p)\rangle =2if_{3M}^{(s)}(pz)^{2}.
\end{equation*}%
In order to extract its value we use the correlation function of non-local
light-ray operators, which enter the definition of the three-particle
distribution amplitude, with corresponding local operator. Such so-called
"non-diagonal" correlation function is given by the following expression
\cite{Ball:2006wn}
\begin{eqnarray}
&&\Pi _{ND}^{s} =i\int d^{4}ye^{-ipy}\langle 0\mid \mathcal{T}\{[\overline{s}%
(z)\sigma _{\mu z}\gamma _{5}gG_{\mu z}(vz)s(0)]  \notag \\
&& \times \lbrack \overline{s}(y)\gamma _{5}s(y)]\} \mid 0\rangle  \notag \\
&&\equiv (pz)^{2}\int D\underline{\alpha }e^{-ipz(\alpha _{2}+v\alpha
_{3})}\pi _{ND}^{s}(\underline{\alpha }).
\end{eqnarray}%
The sum rule for the coupling $f_{3s}$ is derived by expanding the
correlation function in powers of $pz$%
\begin{eqnarray}
\Pi _{ND}^{s} &=&(pz)^{4}\left\{ \Pi _{ND}^{(0)s}+i(pz)\left[ \Pi
_{ND}^{(1A)s}\right. \right.  \notag \\
&&\left. \left. +(2v-1)\Pi _{ND}^{(1B)s}\right] +...\right\} .
\end{eqnarray}%
The hadronic content of the function $\Pi $ has been modeled employing "$%
\eta +\eta ^{\prime }$+continuum" approximation. Then we get the following
sum rule:
\begin{equation*}
f_{3\eta }^{(s)}\frac{h_{\eta }^{(s)}}{m_{s}}e^{-\frac{m_{\eta }^{2}}{M^{2}}%
}+f_{3\eta ^{\prime }}^{(s)}\frac{h_{\eta ^{\prime }}^{(s)}}{m_{s}}e^{-\frac{%
m_{\eta ^{\prime }}^{2}}{M^{2}}}=\mathcal{B}_{M^{2}}\left[ \Pi _{ND}^{(0)s}%
\right] .
\end{equation*}%
The left-hand side of this expression can be modified using information on
mixing of the decay constants:
\begin{eqnarray}
&&\frac{f_{3s}h_{s}}{m_{s}}\left( \sin ^{2}\varphi e^{-\frac{m_{\eta }^{2}}{%
M^{2}}}+\cos ^{2}\varphi e^{-\frac{m_{\eta ^{\prime }}^{2}}{M^{2}}}\right)
\notag \\
&&=\mathcal{B}_{M^2}\left[ \Pi _{ND}^{(0)s}\right].
\end{eqnarray}%
Now having applied the explicit expression for $\mathcal{B}_{M^2}\left[ \Pi
_{ND}^{(0)s}\right]$ we determine $f_{3s}$ using the sum rule:
\begin{eqnarray}
&&\frac{f_{3s}h_{s}}{m_{s}}\left( \sin ^{2}\varphi e^{-\frac{m_{\eta }^{2}}{%
M^{2}}}+\cos ^{2}\varphi e^{-\frac{m_{\eta ^{\prime }}^{2}}{M^{2}}}\right) \nonumber \\
&&=\frac{\alpha _{s}}{73\pi ^{3}}\int_{0}^{s_{0}}dsse^{-\frac{s}{M^{2}}}+%
\frac{1}{12} \langle\frac{\alpha _{s}}{\pi }G^{2}\rangle \nonumber \\
&&-\frac{4\alpha _{s}}{9\pi }m_{s} \langle \overline{s}s\rangle\left[ \frac{%
19}{6}+\gamma _{E}-\ln \frac{M^{2}}{\mu ^{2}}+\int_{s_{0}}^{\infty }\frac{ds%
}{s} e^{-\frac{s}{M^{2}}}\right]  \nonumber \\
&&+\frac{80}{27}\frac{\alpha _{s}\pi }{M^{2}} \langle\overline{s}%
s\rangle^{2}+\frac{1}{ 3M^{2}}m_{s}\langle \overline{s}\sigma gGs\rangle.
\end{eqnarray}
Numerical calculations have been performed at the scale $\mu_{0}=1\,\,%
\mathrm{GeV}$. To evaluate a continuum contribution we set $s_0=1.5\,\,%
\mathrm{GeV}^2$, and varied it within limits $1.3<s_0<1.7\,\, \mathrm{GeV^2}$
to estimate errors. The Borel parameter $M^2$ is changed in the interval $%
0.8<M^2<1.8\,\, \mathrm{GeV}^2$. The parameters have been extracted at $%
M^2=1.3\,\,\mathrm{GeV}^2$. For $f_{3s}(\mu_0)$ we have found:
\begin{equation}
f_{3s}\simeq 0.0041\,\, \mathrm{GeV}^2.
\end{equation}
The varying of $s_0$ in the allowed limits results in errors $\pm 0.00005$,
which may be neglected.

We introduce the parameter $\delta _{M}^{2(s)}$ through the local matrix
element
\begin{align}
\langle 0| \bar s \gamma^\rho ig \widetilde{G}_{\rho\mu} s |M(p)\rangle =
p_\mu f_M^{(s)} \delta^{2(s)}_{M}\,
\end{align}
considering it as the universal one, i.e. we suggest that it does not depend
on the particles $\eta $ and $\eta ^{\prime }$. In the local matrix element
information on the mixing is contained in the decay constants $f_{M}^{(s)}$.
Then we can write
\begin{eqnarray*}
&&f_{s}^{2}\delta _{M}^{4(s)}\left[ \sin ^{2}\varphi e^{-\frac{m_{\eta }^{2}%
}{M^{2}}}+\cos ^{2}\varphi e^{-\frac{m_{\eta ^{\prime }}^{2}}{M^{2}}}\right]=%
\mathcal{B}_{M^2}\left[ \Pi _{0}^{A(s)}\right],  \notag \\
\end{eqnarray*}
where $\mathcal{B}_{M^2}[\Pi _{0}^{A(s)}]$ is given by the expression \cite%
{Ball:2006wn}
\begin{eqnarray}
&&\mathcal{B}_{M^2}[\Pi _{0}^{A(s)}]=\frac{\alpha _{s}}{160\pi ^{3}}%
\int_{0}^{s_{0}}dss^{2}e^{-\frac{s}{M^{2}}}+\frac{1}{72}\langle \frac{\alpha
_{s}}{\pi } G^{2}\rangle  \notag \\
&&\times \int_{0}^{s_{0}}dse^{-\frac{s}{M^{2}}} -\frac{\alpha _{s}}{9\pi }%
m_{s}\langle\overline{s}s\rangle\int_{0}^{s_{0}}dse^{-\frac{s}{M^{2}}}+
\frac{8\pi \alpha _{s}}{9}\langle\overline{s}s\rangle^{2}  \notag \\
&&-\frac{13\alpha _{s}}{54\pi }m_{s}\langle\overline{s}\sigma gGs\rangle +%
\frac{59\pi \alpha _{s}}{81}\frac{m_{0}^{2}}{M^{2}}\langle\overline{s}%
s\rangle^{2}  \notag \\
&&+\frac{\pi }{9M^{2}}\langle\frac{\alpha _{s}}{\pi }G^{2}\rangle
m_{s}\langle \overline{s}s\rangle -\frac{2\alpha _{s}}{\pi }m_{s}\langle
\overline{s}\sigma gGs\rangle  \notag \\
&& \times \left\{ \gamma_{E}-\ln \frac{M^{2}}{\mu ^{2}}+\int_{s_{0}}^{\infty
}\frac{ds}{s} e^{-\frac{s}{M^{2}}}\right\} .
\end{eqnarray}
Computations of $\delta_{M}^{2(s)}$ with the same input parameters as in
previous case, lead to the following prediction:
\begin{equation}
\delta_{M}^{2(s)}(\mu_0)\simeq 0.1896\pm 0.001\,\, \mathrm{GeV}^2.
\end{equation}
As is seen $f_{3s}$ and $\delta_{M}^{2(s)}$ numerically are very close to
the pion's parameters $f_{3\pi}$  and $\delta_{\pi}^2$, respectively.

The values of the quark and quark-gluon condensates at $\mu_0$ utilized in
numerical calculations are listed below:
\begin{eqnarray}
&&\langle\overline{q}q\rangle=(-0.24\pm 0.01)^{3}\ \mathrm{GeV}^{3},\ \
\langle\overline{q}\sigma gGq\rangle=m_{0}^{2}\langle\overline{q}q\rangle,
\notag \\
&&\ m_{0}^{2}=(0.8\pm 0.1)\ \mathrm{GeV}^{2},\, \langle\overline{s}%
s\rangle=[1-(0.2 \pm 0.2]\langle\overline{q}q\rangle,  \notag \\
&& \langle\frac{\alpha _{s}}{\pi }G^{2}\rangle=(0.012\pm 0.006)\ \mathrm{GeV}%
^{4},  \notag \\
&& \langle\overline{s}\sigma gGs\rangle=[1-(0.2\pm 0.2)]\langle\overline{q}%
\sigma gGq\rangle.\
\end{eqnarray}

\end{document}